\documentclass[preprint2]{aastex}
\usepackage{graphicx}
\usepackage{longtable}

\shorttitle{Activity of 50 Long-Period Comets Beyond 5.2 AU}
\shortauthors{S\'arneczky et al.}

\begin{document}

\def\afrho{$Af\rho${} }
\def\afrhomax{$Af\rho_{\rm max}$ }
\def\arcs{$^{\prime\prime}$}

\title{Activity of 50 Long-Period Comets Beyond 5.2 AU}
\author{K. S\'arneczky\altaffilmark{1,2}, 
Gy. M. Szab\'o\altaffilmark{1,2,3}, 
B. Cs\'ak\altaffilmark{1,3}, 
J. Kelemen\altaffilmark{1}, 
G. Marschalk\'o\altaffilmark{4,5}, 
A. P\'al\altaffilmark{1,4}, 
R. Szak\'ats\altaffilmark{1},
T. Szalai\altaffilmark{6}, 
E. Szegedi-Elek\altaffilmark{1}, 
P. Sz\'ekely\altaffilmark{7},
K. Vida\altaffilmark{1},
J. Vink\'o\altaffilmark{1,6,8},
L.L. Kiss\altaffilmark{1,2,9}}

\altaffiltext{1}{Konkoly Observatory, Research Centre for Astronomy and Earth
Sciences, Hungarian Academy of Sciences, H-1121 Budapest, Konkoly Thege
Mikl\'os \'ut 15-17, Hungary}
\altaffiltext{2}{Gothard-Lend\"ulet Research Team, H-9704 Szombathely, Szent
Imre herceg \'ut 112, Hungary}
\altaffiltext{3}{ELTE Gothard  Astrophysical  Observatory, H-9704
Szombathely, Szent Imre herceg \'ut 112, Hungary}
\altaffiltext{4}{E\"otv\"os Lor\'and Tudom\'anyegyetem, H-1117 P\'azm\'any
P\'eter s\'et\'any 1/A, Budapest, Hungary}
\altaffiltext{5}{Baja Observatory of University of Szeged, H-6500 Baja,  Szegedi \'ut III/70, Hungary}
\altaffiltext{6}{Department of Optics \& Quantum Electronics, University of
Szeged, H-6720 Szeged, D\'om t\'er 9, Hungary}
\altaffiltext{7}{Department of Experimental Physics, University of Szeged,
Szeged H-6720, D\'om t\'er 9, Hungary}
\altaffiltext{8}{Department of Astronomy, University of Texas at Austin,
Austin, TX 78712, USA}
\altaffiltext{9}{Sydney Institute for Astronomy, School of Physics A28,
University of Sydney, NSW 2006, Australia}

\begin{abstract}

Remote investigations of the ancient solar system matter has been traditionally carried out through the observations of long-period (LP) comets that are less affected by solar irradiation than the short-period counterparts orbiting much closer to the Sun. Here we summarize the results of our decade-long survey of the distant activity of LP comets. We found that the most important separation in the dataset is based on the dynamical nature of the objects. Dynamically new comets are characterized by a higher level of activity on average: the most active new comets in our sample can be characterized by \afrho{} values $>$3--4 higher than that of our most active returning comets. New comets develop more symmetric comae, suggesting a generally isotropic outflow. Contrary to this, the coma of recurrent comets can be less symmetrical, ocassionally exhibiting negative slope parameters, suggesting sudden variations in matter production. The morphological appearance of the observed comets is rather diverse. A surprisingly large fraction of the comets have long, teniouos tails, but the presence of impressive tails does not show a clear correlation with the brightness of the comets.
\end{abstract}


\maketitle

\section{Introduction}

The origin and behaviour of comets are related to the entire Solar System, its
general history and environment via several aspects. 
It is widely accepted that in the early Solar System a large number of comet-like bodies orbited in the Trans-Neptunian region and beyond, through the Oort-cloud. The group of Trans-Neptunian Objects (TNOs) was recognized as sources for short-period comets and probably various groups of asteroids (e.g. Duncan et al. 2004, Eicher 2013a). The recent exploration of the TNOs by the Herschel space observatory revealed the size and albedos of a handful objects (Lacerda et al. 2014), supporting that comet-like bodies are still present among the TNOs including other objects of different nature (e.g. Fornasier et al. 2013, Duffard et al. 2014). In recent years, the presence of similar comet clouds was suggested in several extrasolar systems, characterized by a prominent infrared excess due to cold debris (e.g. Beichman et al. 2005, Greaves and Wyatt, 2010). These observations show that comets are a common by-products of solar system formation, and they preserve the matter from the outskirts of young solar systems for a long time. Even, the relation of comet dust to ISM relics was suggested in the case of Hale--Bopp (Wooden et al. 2000). 

The observations of cometary activity have been the traditional way of remote investigations. Since the long-period (LP) comets suffered the least solar irradiation, they can be the ideal targets for studying the ancient matter in a fairly intact state.
Thank to the automated telescopes (e.g. Spacewatch, LINEAR, LONEOS, CSS/MLS/SSS, Pan-STARRS), relatively bright, large perihelion-distance
comets are now regularly discovered often several years before the perihelion passage. Hence the study of distant cometary activity has become possible more regularly than before.
Observations at heliocentric distances revealed that comets can be significantly active well beyond the snow line. The sublimation of water ice is excluded in this region because it can only be efficient at a few AU from the Sun, inside indicatively Jupiterâ??s orbit at 5.2 AU (Meech \& Svoren 2004). The main mechanisms that have been proposed to explain cometary activity of comets at large heliocentric distances are: the transition phase between amorphous and crystalline water ice (Prialnik 1992, Capria et al. 2002), the annealing of amorphous water ice (Meech et al. 2009), and the sublimation of more volatile admixtures like CO and/or CO2.

In the past two decades, observational campaigns have been made to reveal the distant activity
of comets, most by with observations of the Jupiter family  (Meech 1991, %
O'Ceallaigh et al. 1995; 
Meech \& Hainaut 1997; %
Lowry et al. 1999; %
Szab\'o et al. 2001; %
Szab\'o et al. 2002; %
Korsun \& Ch{\"o}rny 2003; %
Tozzi et al. 2003; %
Szab\'o et al. 2008; %
Mazzotta Epifani et al. 2009; %
Meech et al. 2009; %
Mazzotta Epifani et al. 2010; %
Korsun et al. 2010; %
Szab\'o et al. 2012; %
Mazzotta Epifani et al. 2014; %
Shi et al. 2014; %
Ivanova et al. 2015). %
However, there is still a lack in the observations of LP comets beyond 5.2 AU. Most importantly, there is a natural deficit of LP comets observed on the inward orbit, because distant comets used to have been discovered near the perihelion (roughly before 2000). For example, as of writing this paper, there are only four comets with well documented observations covering at least 1 AU on the inward orbit beyond 5.2 AU (C/1995 O1 (Fulle at al. 1998), C/2003 WT42 (Korsun et al. 2010), C/2006 S3 (Rousselot et al. 2014),  and C/2012 S1 (Kri\v{s}andov\'a et al. 2014).

The physical mechanism that drives cometary activity is quite different at these large heliocentric distances. The sublimation of water ice is excluded in this region because this process can only be efficient at a few AU from the Sun, inside Jupiterâ??s orbit (Meech \& Svoren 2004). The main mechanisms that have been proposed to explain cometary activity of comets at large heliocentric distances are: the transition phase between amorphous and crystalline water ice (Prialnik 1992, Capria et al. 2002), the annealing of amorphous water ice (Meech et al. 2009), and the sublimation of more volatile admixtures like CO and/or CO2.

According to the recognition by Oort, a dynamically new comet is usually defined as a comet on $a>10,000$ AU orbit, or $P>1$~million yr (Eicher, 2013b, Mazzotta Epifani 2014). These comets represent the early stage of the inward migration (this is why they are called as ``new'' comets). They reside very far from the Sun, and spend the waste majority ($>$99.9\%{}) of their lifetime even outside the heliosphere. Therefore, these comets are exposed to marginal solar irradiation, and only very little modifications by the solar wind. One can expect that the behavior of the dynamically new comets differs from that of the other comets (called as ``returning comets''), which investigation is in the focus of the present paper.

As a result of our decade-long survey about the distant activity of
LP comets, we gained 150 images of 50 comets showing activity beyond the snow line. 
The observations are still on-going with the same instruments, but at the present stage,
we can already answer some important questions related to the activity of these comets.
The observations are presented in the context of the following questions:
\begin{enumerate}
\item{} What is the behaviour of the long-period comets at large heliocentric distances? How does the activity evolve and cease?
\item{} Are the activity profiles similar to each other during the inward and outward orbit?
\item{} What kind of specific correlations can be recognized between the activity parameters ($Af\rho$, slope, tail characteristics), and between these parameters and the ephemerides?
\item{} Can distinct groups be recognized by the activity parameters?
\item{} What can we deduce from short time-scale variations such as outbursts or rapid evolution of matter production?
\end{enumerate}

The paper is structured as follows. The observations and reduction steps are
described in Sect.\ 2, while Sect.\ 3 deals with the detailed observational results. The discussion of the results is given in Sect.\ 4.

\section{Observations}

\begin{table*}
\caption{Cometography. T -- time of perihelion passage; q -- perihelion distance; 1/a -- original semimajor axis; $\Delta$r -- range of r (AU) during the observations. A downward/upward arrow indicates the inward/outward orbit, respectively. All data were taken from the MPC database.}
\begin{center}
\begin{tabular} {lcrcrr}
\hline
Object & T & q (AU) & { } 1/a$_{\rm orig}$ & { } dynamic & { } $\Delta$r { } { } { } { } { } { }\\
\hline
C/1997 BA6 (Spacewatch)  & 1999.11.27. & 3.436  & { } $-$0.000001&  { } new & { } 11.267$\uparrow$ \\
C/1997 J1 (Mueller)      & 1997.05.03. & 2.302  & { } $+$0.001826 &  { } returning   & { } 6.061$\uparrow$ \\
C/1998 W3 (LINEAR)       & 1998.10.06. & 4.915  & { } $+$0.000252 &  { } returning   & { } 6.304$\uparrow$ \\
C/1999 J2 (Skiff)        & 2000.04.06. & 7.110  & { } $+$0.000019 &  { } new         & { } 7.219$\downarrow$ $\rightarrow$ 7.137$\downarrow$ \\
C/2002 V2 (LINEAR)       & 2003.05.13. & 6.812  & { } $+$0.000472 &  { } returning   & { } 6.909$\uparrow$ $\rightarrow$ 6.913$\uparrow$ \\
C/2002 VQ94 (LINEAR)     & 2006.02.08. & 6.784  & { } $+$0.005289 &  { } returning   & { } 8.581$\downarrow$ $\rightarrow$ 11.130$\uparrow$ \\
C/2003 A2 (Gleason)      & 2003.11.06. & 11.426 & { } $+$0.000059 &  { } new         &  { } 11.429$\uparrow$\\
C/2003 K4 (LINEAR)       & 2004.10.13. & 1.024  & { } $+$0.000023 &  { } new         & { } 5.762$\downarrow$ $\rightarrow$ 5.145$\downarrow$ \\
C/2003 WT42 (LINEAR)     & 2006.04.10. & 5.191  & { } $+$0.000051 &  { } new         & { } 7.383$\uparrow$ \\
C/2004 B1 (LINEAR)       & 2006.02.07. & 1.602  & { } $+$0.000036 &  { } new         & { } 9.134$\uparrow$ $\rightarrow$ 10.063$\uparrow$ \\
C/2004 D1 (NEAT)         & 2006.02.10. & 4.975  & { } $+$0.000242 &  { } returning   & { } 7.600$\uparrow$ \\
C/2004 P1 (NEAT)         & 2003.08.08. & 6.014  & { } $+$0.000020 &  { } new         & 7.909$\uparrow$ $\rightarrow$ 7.913$\uparrow$ \\
C/2005 EL173 (LONEOS)    & 2007.03.05. & 3.886  & { } $+$0.000044 &  { } new         & { } 6.944$\downarrow$ $\rightarrow$ 6.705$\uparrow$ \\
C/2005 G1 (LINEAR)       & 2006.02.27. & 4.961  & { } $+$0.000016 &  { } new         & { } 5.566$\downarrow$ \\
C/2005 L3 (McNaught)     & 2008.01.14. & 5.581  & { } $+$0.000061 &  { } new         & 5.620$\uparrow$ $\rightarrow$ 15.152$\uparrow$ \\
C/2005 S4 (McNaught)     & 2007.07.18. & 5.850  & { } $+$0.000409 &  { } returning   & 6.252$\uparrow$ $\rightarrow$ 7.411$\uparrow$ \\
C/2006 A2 (Catalina)     & 2005.05.20. & 5.316  & { } $+$0.000676 &  { } returning   & { } 5.625$\uparrow$ \\
C/2006 K1 (McNaught)     & 2007.07.20. & 4.426  & { } $+$0.000012 &  { } new         & 5.762$\uparrow$ $\rightarrow$ 6.088$\uparrow$ \\
C/2006 K4 (NEAT)         & 2007.11.29. & 3.189  & { } $+$0.000911 &  { } returning   & { } 5.708$\downarrow$ \\
C/2006 M2 (Spacewatch)   & 2005.11.20. & 5.206  & { } unknown$^a$&                 & { } 5.445$\uparrow$ \\
C/2006 S3 (LONEOS)       & 2012.04.16. & 5.131  & { } $+$0.000001 &  { } new         & { } 14.265$\downarrow$ $\rightarrow$ 7.161$\uparrow$ \\
C/2007 B2 (Skiff)        & 2008.08.20. & 2.975  & { } $+$0.001703 &  { } returning   & { } 5.733$\downarrow$ \\
C/2007 D1 (LINEAR)       & 2007.06.19. & 8.794  & { } $+$0.000043 &  { } new         & 9.369$\uparrow$ $\rightarrow$ 10.259$\uparrow$ \\
C/2007 D3 (LINEAR)       & 2007.05.27. & 5.209  & { } $+$0.001307 &  { } returning   & 5.695$\uparrow$ $\rightarrow$ 7.243$\uparrow$ \\
C/2007 G1 (LINEAR)       & 2008.11.16. & 2.647  & { } $+$0.000251 &  { } returning   & { } 5.979$\downarrow$ \\
C/2007 JA21 (LINEAR)     & 2006.11.14. & 5.368  & { } $+$0.000069 &  { } new         & { } 6.506$\uparrow$ $\rightarrow$ 7.143$\uparrow$ \\
C/2007 K1 (Lemmon)       & 2007.05.07. & 9.239  & { } $+$0.002354 &  { } returning   & { } 9.487$\uparrow$ $\rightarrow$ 10.052$\uparrow$ \\
C/2007 M1 (McNaught)     & 2008.08.11. & 7.475  & { } $+$0.000724 &  { } returning   & { } 7.475q $\rightarrow$ 8.551$\uparrow$ \\
C/2007 U1 (LINEAR)       & 2008.08.07. & 3.329  & { } $+$0.000161 &  { } returning   & { } 6.721$\uparrow$ \\
C/2007 VO53 (Spacewatch) & 2010.04.26. & 4.843  & { } $+$0.000091 &  { } new         & { } 6.707$\downarrow$ $\rightarrow$ 5.142$\downarrow$ \\
C/2008 S3 (Boattini)     & 2011.06.07. & 8.018  & { } $+$0.000020 &  { } new         & { } 9.829$\downarrow$ $\rightarrow$ 10.737$\uparrow$ \\
C/2010 D4 (WISE)         & 2009.03.30. & 7.148  & { } $+$0.015990 &  { } 495~yr      &  { } 7.469$\uparrow$ \\
C/2010 G2 (Hill)         & 2011.09.02. & 1.981  & { } $+$0.011105 &  { } 855~yr      & { } 5.457$\downarrow$ \\
C/2010 R1 (LINEAR)       & 2012.05.18. & 5.621  & { } $+$0.000044 &  { } new         & { } 5.639$\downarrow$ $\rightarrow$ 7.253$\uparrow$ \\
C/2010 S1 (LINEAR)       & 2013.05.20. & 5.900  & { } $+$0.000025 &  { } new         & { } 5.908$\uparrow$ $\rightarrow$ 6.927$\uparrow$ \\
C/2010 U3 (Boattini)     & 2019.02.26. & 8.446  & { } $+$0.000058 &  { } new         & { } 13.563$\downarrow$ $\rightarrow$ 12.381$\downarrow$ \\
C/2011 F1 (LINEAR)       & 2013.01.08. & 1.819  & { } $+$0.000006 &  { } new         & { } 6.926$\downarrow$ \\
C/2011 KP36 (Spacewatch) & 2016.05.26. & 4.883  & { } $+$0.025812 &  { } 241~yr      & { } 8.478$\downarrow$ $\rightarrow$ 8.212$\downarrow$ \\
C/2012 K1 (PANSTARRS)    & 2014.08.27. & 1.055  & { } $+$0.000035 &  { } new         & { } 8.784$\downarrow$ $\rightarrow$ 5.210$\downarrow$ \\
C/2012 K8 (Lemmon)       & 2014.08.19. & 6.463  & { } $+$0.000041 &  { } new         & { } 6.826$\downarrow$ $\rightarrow$ 6.478$\uparrow$ \\
C/2012 LP26 (Palomar)    & 2015.08.16. & 6.536  & { } $+$0.000039 &  { } new         & { } 8.172$\downarrow$ $\rightarrow$ 6.836$\downarrow$ \\
C/2012 S1 (ISON)         & 2013.11.28. & 0.013  & { } $+$0.000006 &  { } new         & { } 5.997$\downarrow$ $\rightarrow$ 5.188$\downarrow$ \\
C/2012 U1 (PANSTARRS)    & 2014.07.04. & 5.264  & { } $+$0.001277 &  { } returning   & { } 6.609$\downarrow$ $\rightarrow$ 5.278$\downarrow$ \\
C/2013 G9 (Tenagra)      & 2015.01.14. & 5.338  & { } $+$0.000072 &  { } new         & { } 6.822$\downarrow$ $\rightarrow$ 5.799$\downarrow$ \\
C/2013 L2 (Catalina)     & 2012.05.11. & 4.873  & { } $+$0.000076 &  { } new         & { } 5.731$\uparrow$ $\rightarrow$ 6.003$\uparrow$ \\
C/2013 P3 (Palomar)      & 2014.11.23. & 8.647  & { } $+$0.000042 &  { } new         & { } 9.059$\downarrow$ $\rightarrow$ 8.648$\downarrow$ \\
C/2013 X1 (PANSTARRS)    & 2016.04.20. & 1.314  & { } $+$0.000276 &  { } returning   & { } 8.712$\downarrow$ $\rightarrow$ 6.255$\downarrow$ \\
C/2014 L5 (Lemmon)       & 2014.11.26. & 6.203  & { } $+$0.000208 &  { } returning   & { } 6.207$\downarrow$ \\
C/2014 M2 (Christensen)  & 2014.07.18. & 6.909  & { } $+$0.000796 &  { } returning   & { } 6.925$\uparrow$ \\
C/2014 R3 (PANSTARRS)    & 2016.08.07. & 7.274  & { } $+$0.000130 &  { } returning   & { } 8.344$\downarrow$ \\

\hline
\end{tabular}
$^a$ Only parabolic orbital solution exists.
\end{center}
\end{table*}

\begin{table*}
\caption{Observation summary. R -- heliocentric distance; $\Delta$ -- geocentric distance; $\alpha$ - solar phase angle, PA$_{dust}$ -- predicted position angle of dust tail. All data were taken from the NASA/Horizons service (http://ssd.jpl.nasa.gov/horizons.cgi). I/O: inboud/outbound (pre-/post-perihelion) leg of the orbit.} 
\begin{center}
\begin{tabular} {lcrrrrrrc}
\hline
Object & obs. time (UT) & { }R (AU) & { }$\Delta$ (AU) & { }{ } $\alpha$ & { } $PA_{dust}$ & { } Exp. (s) & I/O\\
\hline
C/1997 BA6 (Spacewatch)  &  2003.09.19. 18:49  &  11.267 & 10.563 & { }{ } 3.8  & { } 183.2  & { } 4$\times$280 & O \\
C/1997 J1 (Mueller)      &  1998.11.25. 00:03  &   6.061 &  5.215 & { }{ } 5.2  & { }  39.6  & { } 3$\times$240 & O \\
C/1998 W3 (LINEAR)       &  2000.03.10. 18:35  &   6.304 &  6.474 & { }{ } 8.8  & { }  91.0  & { } 3$\times$300 & O \\
C/1999 J2 (Skiff)        &  1999.09.24. 18:21  &   7.219 &  7.582 & { }{ } 7.2  & { }  10.7  & { } 2$\times$300 & I \\
                         &  2000.01.01. 03:52  &   7.137 &  7.473 & { }{ } 7.3  & { }  18.3  & { } 600 & I \\
C/2002 V2 (LINEAR)       &  2003.11.05. 00:18  &   6.909 &  5.940 & { }{ } 1.9  & { }  87.0  & { } 3$\times$180 & O \\
                         &  2003.11.09. 00:55  &   6.913 &  5.935 & { }{ } 1.4  & { }  86.9  & { } 3$\times$200 & O \\
C/2002 VQ94 (LINEAR)     &  2003.11.03. 17:59  &   8.581 &  7.892 & { }{ } 5.0  & { } 178.2  & { } 4$\times$180 & I \\
                         &  2008.02.25. 03:40  &   8.296 &  7.758 & { }{ } 5.9  & { }   4.5  & { } 12$\times$120 & O \\
                         &  2008.03.31. 01:58  &   8.421 &  7.565 & { }{ } 3.7  & { }   2.2  & { } 8$\times$180 & O \\
                         &  2008.12.29. 04:53  &   9.482 &  9.884 & { }{ } 5.3  & { }   3.3  & { } 4$\times$180 & O \\
                         &  2009.04.02. 01:56  &   9.874 &  8.951 & { }{ } 2.4  & { }   1.6  & { } 5$\times$180 & O \\
                         &  2009.04.16. 22:07  &   9.936 &  8.953 & { }{ } 1.2  & { }   0.6  & { } 6$\times$180 & O \\
                         &  2010.01.15. 04:46  &  11.130 & 11.318 & { }{ } 4.9  & { }   2.9  & { } 10$\times$180 & O \\
C/2003 A2 (Gleason)      &  2003.12.27. 02:16  &  11.429 & 10.801 & { }{ } 3.9  & { } 281.8  & { } 4$\times$260 & O \\
C/2003 K4 (LINEAR)       &  2003.07.06. 00:45  &   5.762 &  5.202 & { }{ } 8.9  & { } 113.6  & { } 3$\times$120 & I \\
                         &  2003.09.09. 19:21  &   5.145 &  4.588 & { }{ } 9.9  & { }  93.0  & { } 5$\times$120 & I \\
C/2003 WT42 (LINEAR)     &  2008.03.31. 00:55  &   7.383 &  6.468 & { }{ } 3.3  & { } 321.7  & { } 9$\times$180 & O \\
C/2004 B1 (LINEAR)       &  2008.08.22. 19:35  &   9.134 &  9.600 & { }{ } 5.5  & { } 127.6  & { } 9$\times$200 & O \\
                         &  2008.08.24. 19:16  &   9.148 &  9.618 & { }{ } 5.5  & { } 127.9  & { } 9$\times$200 & O \\
                         &  2008.12.30. 04:54  &  10.063 &  9.885 & { }{ } 5.6  & { } 128.8  & { } 9$\times$180 & O \\
C/2004 D1 (NEAT)         &  2008.03.31. 02:25  &   7.600 &  6.807 & { }{ } 4.8  & { } 333.2  & { } 8$\times$180 & O \\
C/2004 P1 (NEAT)         &  2005.08.29. 21:56  &   7.909 &  7.096 & { }{ } 4.6  & { } 238.6  & { } 5$\times$230 & O \\
                         &  2005.08.30. 23:27  &   7.913 &  7.095 & { }{ } 4.5  & { } 238.6  & { } 4$\times$240 & O \\
C/2005 EL173 (LONEOS)    &  2005.03.31. 21:44  &   6.944 &  6.056 & { }{ } 4.6  & { }  65.7  & { } 4$\times$210 & I \\
                         &  2008.12.29. 16:53  &   6.705 &  6.965 & { }{ } 4.1  & { } 121.4  & { } 8$\times$180 & O \\
C/2005 G1 (LINEAR)       &  2005.04.03. 01:52  &   5.566 &  5.432 & { }{ } 10.3 & { } 152.5  & { } 3$\times$180 & I \\
C/2005 L3 (McNaught)     &  2008.03.31. 02:52  &   5.620 &  5.154 & { }{ } 9.4  & { } 129.2  & { } 8$\times$180 & O \\
                         &  2008.05.28. 23:11  &   5.677 &  4.903 & { }{ } 7.1  & { } 125.0  & { } 9$\times$120 & O \\
                         &  2008.12.31. 05:14  &   6.136 &  6.461 & { }{ } 8.5  & { } 122.8  & { } 4$\times$120 & O \\
                         &  2009.04.18. 02:20  &   6.486 &  5.774 & { }{ } 6.6  & { } 113.6  & { } 8$\times$150 & O \\
                         &  2009.05.01. 19:41  &   6.535 &  5.887 & { }{ } 7.1  & { } 111.3  & { } 4$\times$280 & O \\
                         &  2009.12.29. 04:47  &   7.525 &  7.429 & { }{ } 7.5  & { } 109.8  & { } 9$\times$140 & O \\
                         &  2010.02.22. 04:01  &   7.776 &  7.060 & { }{ } 5.3  & { } 103.9  & { } 9$\times$120 & O \\
                         &  2010.05.23. 20:12  &   8.203 &  8.068 & { }{ } 7.1  & { }  91.3  & { } 8$\times$120 & O \\
                         &  2010.06.12. 21:17  &   8.299 &  8.454 & { }{ } 6.9  & { }  91.2  & { } 11$\times$180 & O \\
                         &  2014.02.24. 00:11  &  15.152 & 14.303 & { }{ } 2.0  & { }  67.7  & { } 12$\times$180 & O \\
C/2005 S4 (McNaught)     &  2008.05.26. 00:43  &   6.252 &  5.922 & { }{ } 9.0  & { } 139.4  & { } 12$\times$120 & O \\
                         &  2008.08.20. 20:55  &   6.490 &  6.241 & { }{ } 8.8  & { } 127.5  & { } 12$\times$180 & O \\
                         &  2008.08.22. 20:14  &   6.495 &  6.261 & { }{ } 8.8  & { } 127.4  & { } 12$\times$150 & O \\
                         &  2008.08.26. 21:07  &   6.508 &  6.305 & { }{ } 8.9  & { } 127.3  & { } 5$\times$180 & O \\
                         &  2009.04.02. 02:47  &   7.282 &  7.161 & { }{ } 7.9  & { } 127.5  & { } 6$\times$180 & O \\
                         &  2009.04.20. 01:47  &   7.355 &  7.152 & { }{ } 7.8  & { } 124.7  & { } 5$\times$180 & O \\
                         &  2009.05.03. 01:10  &   7.411 &  7.172 & { }{ } 7.7  & { } 121.9  & { } 6$\times$180 & O \\
C/2006 A2 (Catalina)     &  2006.01.22. 19:14  &   5.625 &  5.192 & { }{ } 9.4  & { }  41.1  & { } 9$\times$220 & O \\
\hline
\end{tabular}
\end{center}
\end{table*}

\setcounter{table}{1}
\begin{table*}
\caption{cont.}
\begin{center}
\begin{tabular} {lcrrrrrrc}
\hline
Object & obs. time (UT) & { }R (AU) & { }$\Delta$ (AU) & { }{ } $\alpha$ & { } $PA_{dust}$ & { } exp. & I/O\\
\hline
C/2006 K1 (McNaught)     &  2008.10.25. 02:14  &   5.762 &  5.003 & { }{ } 6.9  & { } 206.1   & { } 9$\times$180 & O \\
                         &  2008.11.28. 02:22  &   5.931 &  4.949 & { }{ } 1.0  & { } 207.7   & { } 9$\times$180 & O \\
                         &  2008.11.28. 19:30  &   5.935 &  4.951 & { }{ } 0.9  & { } 207.8   & { } 9$\times$180 & O \\
                         &  2008.12.28. 21:03  &   6.088 &  5.219 & { }{ } 4.7  & { } 208.8   & { } 9$\times$180 & O \\
C/2006 K4 (NEAT)         &  2006.06.20. 23:47  &   5.708 &  5.061 & { }{ } 8.4  & { }   5.8   & { } 5$\times$240 & I \\
C/2006 M2 (Spacewatch)   &  2006.06.20. 22:56  &   5.445 &  4.608 & { }{ } 6.6  & { }  75.4   & { } 7$\times$220 & O \\
C/2006 S3 (LONEOS)       &  2006.09.23. 23:32  &  14.265 & 13.275 & { }{ } 0.7  & { }  78.9   & { } 9$\times$120 & I \\
                         &  2008.08.24. 22:10  &  10.569 &  9.634 & { }{ } 2.2  & { }  77.7   & { } 12$\times$180 & I \\
                         &  2008.08.25. 21:55  &  10.563 &  9.623 & { }{ } 2.1  & { }  77.7   & { } 12$\times$150 & I \\
                         &  2008.08.26. 21:34  &  10.558 &  9.613 & { }{ } 2.0  & { }  77.7   & { } 12$\times$140 & I \\
                         &  2008.11.28. 17:48  &  10.055 &  9.915 & { }{ } 5.6  & { }  74.9   & { } 8$\times$180 & I \\
                         &  2008.12.28. 17:18  &   9.894 & 10.271 & { }{ } 5.2  & { }  74.5   & { } 9$\times$180 & I \\
                         &  2009.09.15. 21:54  &   8.504 &  7.568 & { }{ } 2.6  & { }  76.0   & { } 3$\times$180 & I \\
                         &  2009.12.26. 16:37  &   7.975 &  8.534 & { }{ } 5.6  & { }  74.1   & { } 12$\times$120 & I \\
                         &  2010.06.12. 00:44  &   7.137 &  6.732 & { }{ } 7.7  & { }  76.1   & { } 12$\times$180 & I \\
                         &  2014.02.24. 01:56  &   7.161 &  6.432 & { }{ } 5.7  & { } 100.3   & { } 7$\times$150 & O \\
C/2007 B2 (Skiff)        &  2007.03.06. 23:26  &   5.733 &  4.985 & { }{ } 7.0  & { } 298.7   & { } 12$\times$120 & I \\
C/2007 D1 (LINEAR)       &  2008.12.30. 04:19  &   9.369 &  9.233 & { }{ } 6.0  & { } 251.7   & { } 9$\times$180 & O\\
                         &  2009.04.16. 21:41  &   9.597 &  8.713 & { }{ } 3.0  & { } 253.3   & { } 9$\times$180 & O \\
                         &  2009.04.17. 23:26  &   9.600 &  8.724 & { }{ } 3.1  & { } 253.3   & { } 12$\times$150 & O \\
                         &  2009.04.19. 21:50  &   9.604 &  8.745 & { }{ } 3.3  & { } 253.4   & { } 12$\times$150 & O \\
                         &  2009.12.29. 03:43  &  10.259 & 10.349 & { }{ } 5.4  & { } 253.9   & { } 12$\times$180 & O \\
C/2007 D3 (LINEAR)       &  2008.03.31. 20:18  &   5.695 &  5.198 & { }{ } 9.1  & { } 242.6   & { } 8$\times$180 & O \\
                         &  2008.12.24. 02:56  &   6.706 &  6.270 & { }{ } 7.8  & { } 243.9   & { } 9$\times$180 & O \\
                         &  2009.04.16. 19:30  &   7.229 &  6.626 & { }{ } 6.7  & { } 248.5   & { } 8$\times$180 & O \\
                         &  2009.04.18. 21:18  &   7.239 &  6.664 & { }{ } 6.8  & { } 248.5   & { } 12$\times$150 & O \\
                         &  2009.04.19. 21:15  &   7.243 &  6.682 & { }{ } 6.9  & { } 248.5   & { } 12$\times$150 & O \\
C/2007 G1 (LINEAR)       &  2007.05.03. 00:52  &   5.979 &  5.544 & { }{ } 9.1  & { } 359.2   & { } 12$\times$120 & I \\
C/2007 JA21 (LINEAR)     &  2008.04.01. 01:17  &   6.506 &  6.098 & { }{ } 8.3  & { }  15.2   & { } 8$\times$180 & O \\
                         &  2008.08.27. 19:18  &   7.143 &  7.203 & { }{ } 8.1  & { }   4.0   & { } 10$\times$150 & O \\
C/2007 K1 (Lemmon)       &  2008.05.28. 00:13  &   9.487 &  9.084 & { }{ } 5.7  & { } 144.4   & { } 8$\times$180 & O \\
                         &  2008.08.20. 20:12  &   9.605 &  9.604 & { }{ } 6.0  & { } 136.7   & { } 12$\times$180 & O \\
                         &  2008.08.24. 19:49  &   9.611 &  9.640 & { }{ } 6.0  & { } 136.6   & { } 9$\times$180 & O \\
                         &  2009.04.20. 01:19  &  10.052 &  9.775 & { }{ } 5.6  & { } 137.2   & { } 9$\times$180 & O \\
C/2007 M1 (McNaught)     &  2008.08.22. 21:07  &   7.475 &  6.958 & { }{ } 6.9  & { } 111.3   & { } 7$\times$150 & O \\
                         &  2008.08.24. 20:20  &   7.475 &  6.984 & { }{ } 7.0  & { } 111.3   & { } 9$\times$180 &  O\\
                         &  2009.05.04. 01:42  &   7.657 &  7.072 & { }{ } 6.4  & { } 110.1   & { } 4$\times$150 & O \\
                         &  2010.06.12. 23:00  &   8.551 &  7.919 & { }{ } 5.5  & { }  98.9   & { } 9$\times$150 & O \\
C/2007 U1 (LINEAR)       &  2010.06.13. 00:23  &   6.721 &  6.266 & { }{ } 8.0  & { }  80.3   & { } 12$\times$180 & O \\
C/2007 VO53 (Spacewatch) &  2008.08.21. 01:02  &   6.707 &  6.819 & { }{ } 8.5  & { } 180.2   & { } 12$\times$150 & I \\
                         &  2008.10.26. 02:18  &   6.387 &  5.566 & { }{ } 5.4  & { } 174.8   & { } 2$\times$150 & I \\
                         &  2008.12.23. 20:46  &   6.115 &  5.363 & { }{ } 6.3  & { } 160.4   & { } 6$\times$180 & I \\
                         &  2009.09.16. 01:12  &   5.142 &  5.002 & { }{ } 11.3 & { } 186.0   & { } 5$\times$180 & I \\
\hline                                                                                            
\end{tabular}                                                                                     
\end{center}                                                                                      
\end{table*}                                                                                      
                                                                                                  
\setcounter{table}{1}
\begin{table*}
\caption{cont.}
\begin{center}
\begin{tabular} {lcrrrrrrc}
\hline
Object & obs. time (UT) & { }R (AU) & { }$\Delta$ (AU) & { }{ } $\alpha$ & { } $PA_{dust}$ & { } exp. & I/O\\
\hline
C/2008 S3 (Boattini)     &  2008.10.25. 01:43  &   9.829 &  9.018 & { }{ } 3.5  & { }  98.5   & { } 9$\times$180 & I \\
                         &  2008.12.23. 21:21  &   9.633 &  8.789 & { }{ } 3.1  & { }  94.7   & { } 9$\times$180 & I \\
                         &  2008.12.28. 19:54  &   9.617 &  8.825 & { }{ } 3.6  & { }  94.5   & { } 12$\times$160 & I \\
                         &  2008.12.30. 19:31  &   9.611 &  8.842 & { }{ } 3.8  & { }  94.4   & { } 12$\times$150 & I \\
                         &  2008.12.31. 19:31  &   9.608 &  8.850 & { }{ } 3.9  & { }  94.3   & { } 15$\times$150 & I \\
                         &  2009.09.15. 02:11  &   8.867 &  8.395 & { }{ } 5.9  & { }  94.8   & { } 8$\times$180 & I \\
                         &  2013.09.27. 18:39  &   9.470 &  8.716 & { }{ } 4.2  & { }  70.2   & { } 7$\times$120 & O \\
                         &  2014.07.24. 21:53  &  10.489 &  9.558 & { }{ } 2.3  & { }  69.9   & { } 4$\times$150 & O \\
                         &  2014.07.25. 22:26  &  10.493 &  9.557 & { }{ } 2.3  & { }  69.9   & { } 4$\times$150 & O \\
                         &  2014.09.29. 20:01  &  10.737 & 10.185 & { }{ } 4.6  & { }  69.9   & { } 6$\times$150 & O \\
C/2010 D4 (WISE)         &  2010.03.24. 03:24  &   7.469 &  7.242 & { }{ } 7.6  & { } 167.5   & { } 8$\times$150 & O \\
C/2010 G2 (Hill)         &  2010.05.01. 21:38  &   5.457 &  4.656 & { }{ } 7.0  & { } 183.2   & { } 9$\times$140 & I \\
C/2010 R1 (LINEAR)       &  2012.03.18. 03:36  &   5.639 &  5.200 & { }{ } 9.5  & { }  96.5   & { } 7$\times$60 & I \\
                         &  2014.02.24. 00:49  &   7.253 &  6.314 & { }{ } 2.6  & { }  91.1   & { } 9$\times$150 & O \\
C/2010 S1 (LINEAR)       &  2013.07.03. 21:56  &   5.908 &  5.349 & { }{ } 8.7  & { }  36.6   & { } 3$\times$120 & O \\
                         &  2013.10.02. 18:53  &   5.977 &  5.696 & { }{ } 9.4  & { }  28.1   & { } 3$\times$90 & O \\
                         &  2014.03.15. 03:38  &   6.265 &  6.770 & { }{ } 7.6  & { }  33.0   & { } 3$\times$120 & O \\
                         &  2014.10.21. 16:59  &   6.927 &  7.245 & { }{ } 7.6  & { }  29.8   & { } 5$\times$60 & O \\
C/2010 U3 (Boattini)     &  2013.12.27. 19:33  &  13.563 & 12.910 & { }{ } 3.2  & { } 194.5   & { } 9$\times$150 & I \\
                         &  2014.07.27. 01:03  &  12.715 & 13.076 & { }{ } 4.2  & { } 204.5   & { } 5$\times$150 & I \\
                         &  2014.10.20. 02:03  &  12.381 & 11.547 & { }{ } 2.6  & { } 201.9   & { } 3$\times$150 & I \\
C/2011 F1 (LINEAR)       &  2011.03.25. 21:13  &   6.926 &  6.510 & { }{ } 7.7  & { } 275.9   & { } 13$\times$150 & I \\
C/2011 KP36 (Spacewatch) &  2013.08.05. 21:03  &   8.478 &  7.821 & { }{ } 5.4  & { } 271.1   & { } 2$\times$150 & I \\
                         &  2013.09.28. 18:03  &   8.212 &  8.332 & { }{ } 6.9  & { } 273.0   & { } 4$\times$90 & I \\
C/2012 K1 (PANSTARRS)    &  2012.05.20. 01:21  &   8.784 &  8.014 & { }{ } 4.5  & { } 125.9   & { } 5$\times$150 & I \\
                         &  2013.03.12. 03:38  &   6.352 &  6.280 & { }{ } 9.0  & { } 124.3   & { } 3$\times$150 & I \\
                         &  2013.07.14. 22:21  &   5.210 &  4.739 & { }{ } 10.4 & { } 112.2   & { } 9$\times$150 & I \\
C/2012 K8 (Lemmon)       &  2013.09.27. 18:16  &   6.826 &  6.658 & { }{ } 8.4  & { } 140.3   & { } 5$\times$120 & I \\
                         &  2014.06.03. 22:50  &   6.484 &  6.147 & { }{ } 8.7  & { } 141.0   & { } 7$\times$150 & I \\
                         &  2014.07.25. 20:55  &   6.465 &  6.213 & { }{ } 8.9  & { } 128.4   & { } 7$\times$150 & I \\
                         &  2014.10.21. 17:54  &   6.477 &  6.772 & { }{ } 8.2  & { } 122.3   & { } 7$\times$120 & O \\
C/2012 LP26 (Palomar)    &  2013.08.01. 20:47  &   8.172 &  7.816 & { }{ } 6.8  & { } 278.9   & { } 12$\times$120 & I \\
                         &  2014.06.03. 23:34  &   7.162 &  6.276 & { }{ } 4.2  & { } 279.8   & { } 3$\times$150 & I \\
                         &  2014.07.26. 22:48  &   7.026 &  6.274 & { }{ } 5.9  & { } 280.1   & { } 9$\times$150 & I \\
                         &  2014.10.21. 17:21  &   6.836 &  7.221 & { }{ } 7.5  & { } 281.8   & { } 5$\times$120 & I \\
C/2012 S1 (ISON)         &  2012.10.21. 03:52  &   5.997 &  5.965 & { }{ } 9.5  & { } 296.1   & { } 5$\times$120 & I \\
                         &  2012.11.14. 03:34  &   5.639 &  5.209 & { }{ } 9.4  & { } 293.4   & { } 6$\times$200 & I \\
                         &  2013.01.07. 20:21  &   5.188 &  4.219 & { }{ } 2.0  & { } 296.7   & { } 5$\times$120 & I \\
C/2012 U1 (PANSTARRS)    &  2013.01.05. 18:28  &   6.609 &  6.337 & { }{ } 8.4  & { } 186.0   & { } 4$\times$150 & I \\
                         &  2013.01.07. 17:43  &   6.601 &  6.360 & { }{ } 8.4  & { } 185.9   & { } 3$\times$150 & I \\
                         &  2013.07.12. 00:19  &   5.895 &  6.369 & { }{ } 8.4  & { } 206.3   & { } 4$\times$150 & I \\
                         &  2013.07.20. 01:03  &   5.869 &  6.244 & { }{ } 9.0  & { } 207.2   & { } 4$\times$120 & I \\
                         &  2014.08.25. 02:24  &   5.278 &  5.668 & { }{ } 9.8  & { } 278.9   & { } 3$\times$150 & O \\
C/2013 G9 (Tenagra)      &  2013.06.09. 22:47  &   6.822 &  5.906 & { }{ } 3.9  & { }  73.0   & { } 7$\times$120 & I \\
                         &  2013.07.02. 22:39  &   6.723 &  6.019 & { }{ } 6.6  & { }  73.6   & { } 3$\times$240 & I \\
                         &  2014.03.14. 03:20  &   5.799 &  5.203 & { }{ } 8.3  & { }  70.7   & { } 6$\times$150 & I \\
\hline                                                                                            
\end{tabular}                                                                                     
\end{center}
\end{table*}

\setcounter{table}{1}
\begin{table*}
\caption{cont.}
\begin{center}
\begin{tabular} {lcrrrrrrc}
\hline
Object & obs. time (UT) & { }R (AU) & { }$\Delta$ (AU) & { }{ } $\alpha$ & { } $PA_{dust}$ & { } exp. & I/O\\
\hline
C/2013 L2 (Catalina)     &  2013.06.08. 00:15  &   5.731 &  5.078 & { }{ } 8.3  & { } 155.2   & { } 9$\times$120 & O \\
                         &  2013.06.12. 22:33  &   5.750 &  5.092 & { }{ } 8.3  & { } 155.0   & { } 5$\times$150 & O \\
                         &  2013.07.11. 23:27  &   5.868 &  5.299 & { }{ } 8.7  & { } 154.2   & { } 5$\times$150 & O \\
                         &  2013.08.12. 20:44  &   6.003 &  5.714 & { }{ } 9.5  & { } 154.1   & { } 5$\times$180 & O \\
C/2013 P3 (Palomar)      &  2013.08.16. 21:36  &   9.059 &  8.355 & { }{ } 4.8  & { } 338.7   & { } 15$\times$100 & I \\
                         &  2013.09.27. 18:59  &   8.990 &  8.100 & { }{ } 3.1  & { } 338.1   & { } 7$\times$120 & I \\
                         &  2013.10.31. 19:13  &   8.938 &  8.264 & { }{ } 4.9  & { } 338.0   & { } 6$\times$150 & I \\
                         &  2013.12.04. 16:56  &   8.890 &  8.685 & { }{ } 6.3  & { } 338.3   & { } 5$\times$150 & I \\
                         &  2014.07.25. 00:11  &   8.676 &  8.162 & { }{ } 6.0  & { } 340.8   & { } 9$\times$150 & I \\
                         &  2014.10.25. 19:53  &   8.648 &  7.852 & { }{ } 4.1  & { } 340.4   & { } 7$\times$150 & I \\
C/2013 X1 (PANSTARRS)    &  2013.12.27. 20:08  &   8.712 &  7.746 & { }{ } 1.3  & { } 113.5   & { } 7$\times$150 & I\\
                         &  2014.03.14. 20:27  &   8.119 &  7.903 & { }{ } 6.9  & { } 102.2   & { } 9$\times$120 & I \\
                         &  2014.08.23. 02:08  &   6.817 &  7.447 & { }{ } 6.4  & { } 112.2   & { } 3$\times$150 & I \\
                         &  2014.08.25. 02:03  &   6.801 &  7.404 & { }{ } 6.6  & { } 112.5   & { } 7$\times$150 & I \\
                         &  2014.10.28. 03:37  &   6.255 &  5.809 & { }{ } 8.5  & { } 118.1   & { } 5$\times$120 & I \\
C/2014 L5 (Lemmon)       &  2014.10.27. 19:38  &   6.207 &  5.510 & { }{ } 6.9  & { } 105.1   & { } 9$\times$150 & I \\
C/2014 M2 (Christensen)  &  2014.09.29. 18:28  &   6.925 &  6.829 & { }{ } 8.3  & { } 239.7   & { } 11$\times$150 & O \\
C/2014 R3 (PANSTARRS)    &  2014.10.28. 16:54  &   8.344 &  8.261 & { }{ } 6.8  & { } 129.9   & { } 9$\times$150 & I \\
\hline
\end{tabular}
\end{center}
\end{table*}

\begin{figure*}
\begin{tabular}{ccc}
\small C/2002 VQ94 [2008--03--30] &\small  C/2003 WT42 [2008--03--30] &\small  C/2005 L3 [2008-04-17]\\
\includegraphics[height=4.64cm]{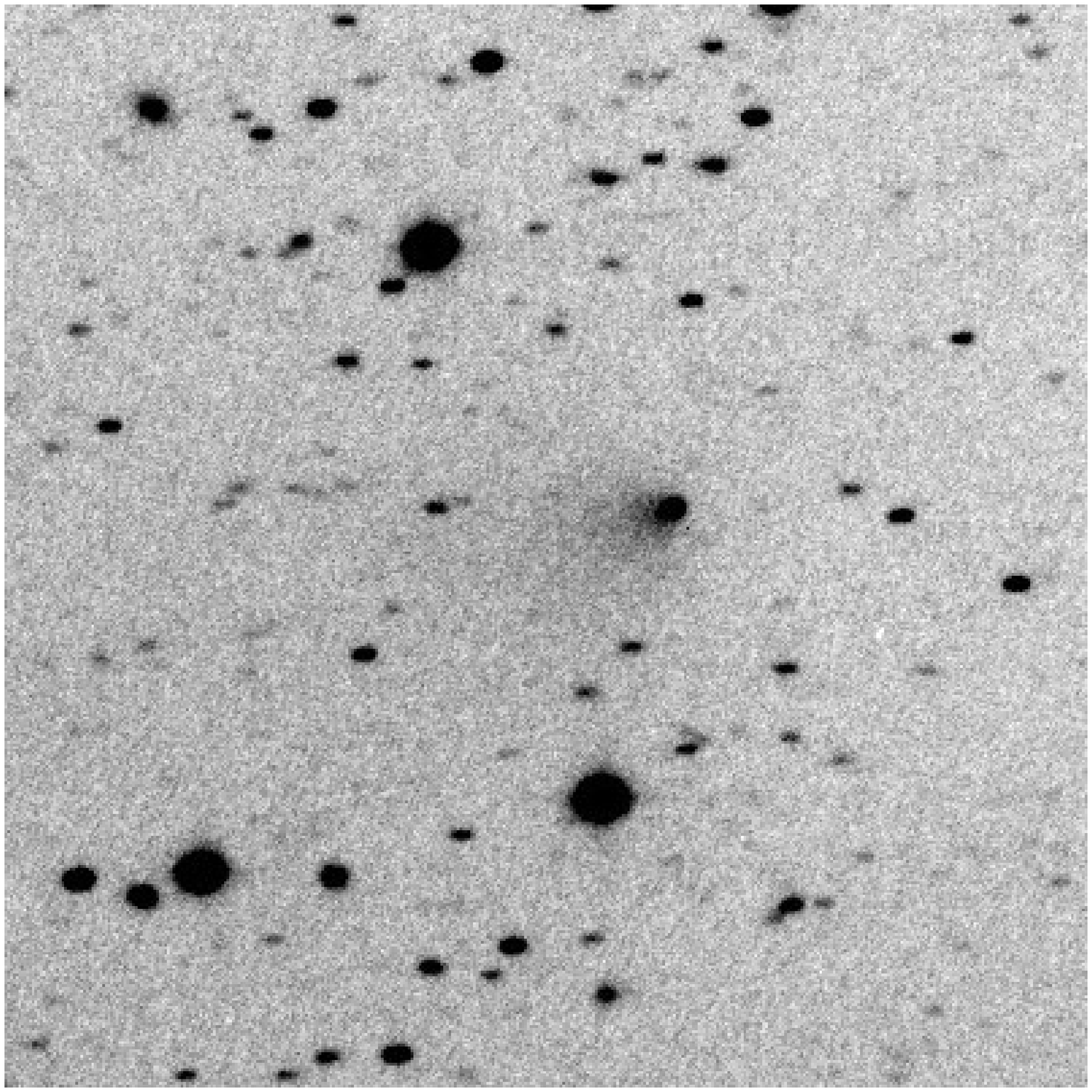}&
\includegraphics[height=4.64cm]{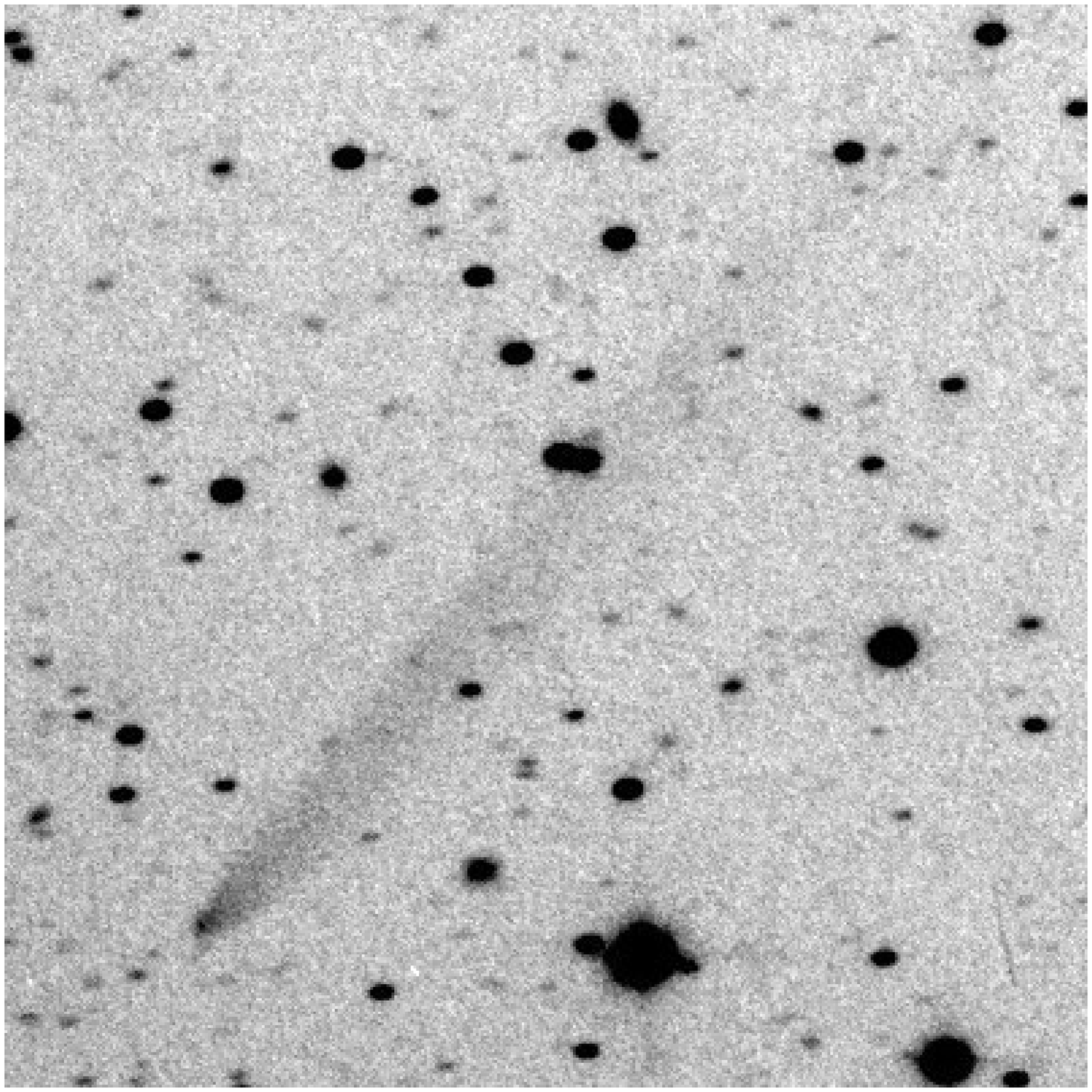}&
\noindent\includegraphics[height=4.64cm]{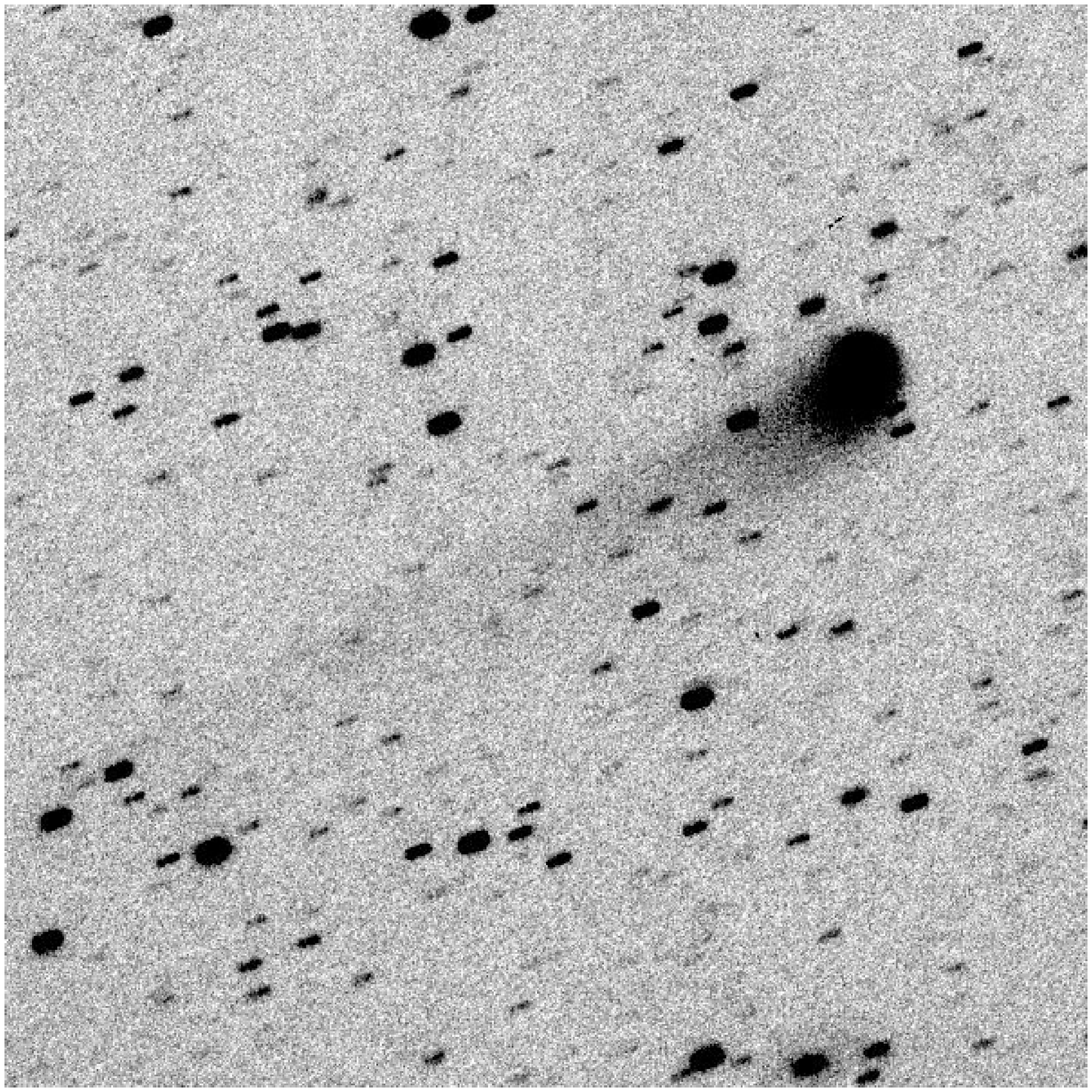}\\
\small C/2006 K1 [2008--12--28] & \small C/2006 S3 [2010--06--12] & \small C/2006 S3 [2009--04--17]\\
\includegraphics[height=4.64cm]{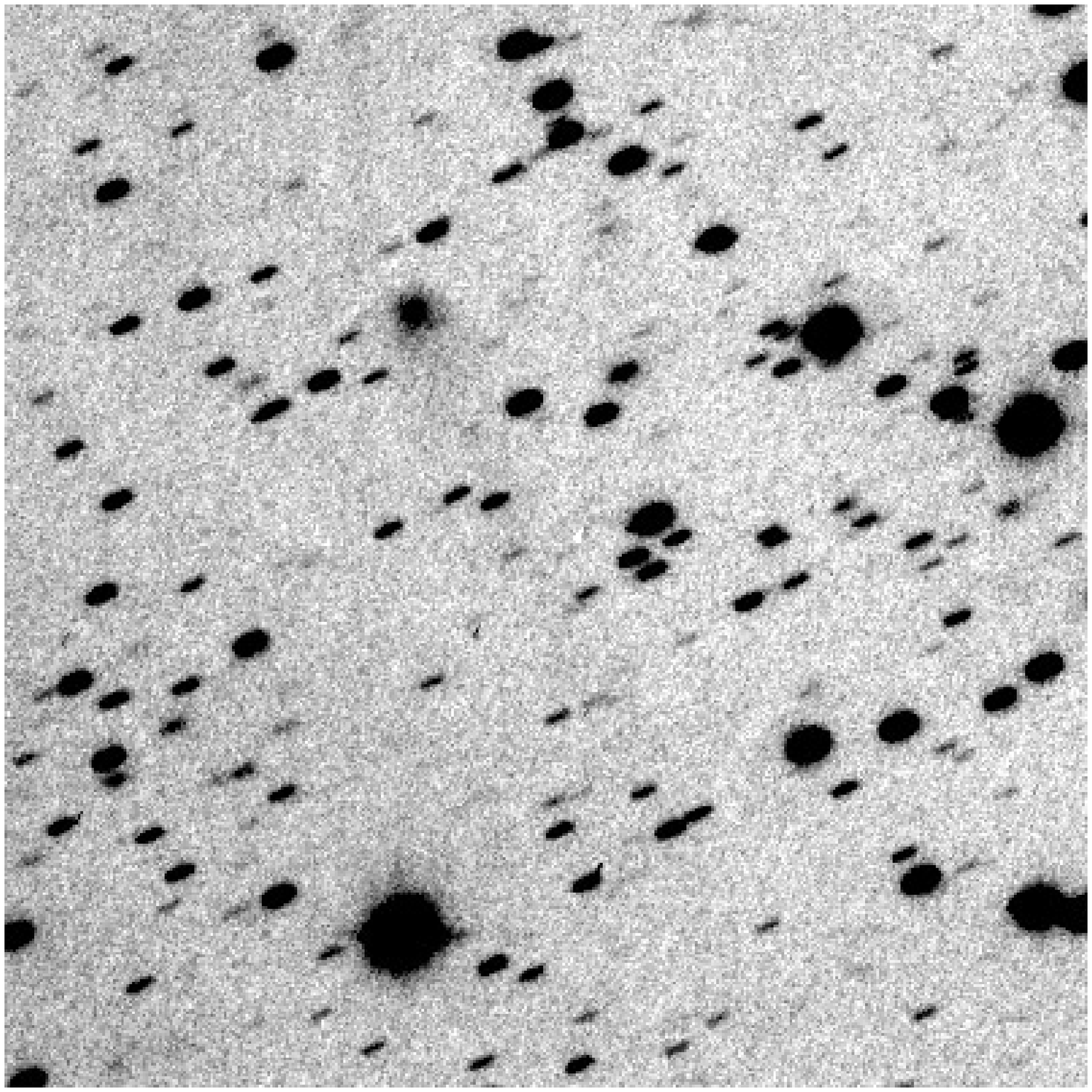}&
\noindent\includegraphics[height=4.64cm]{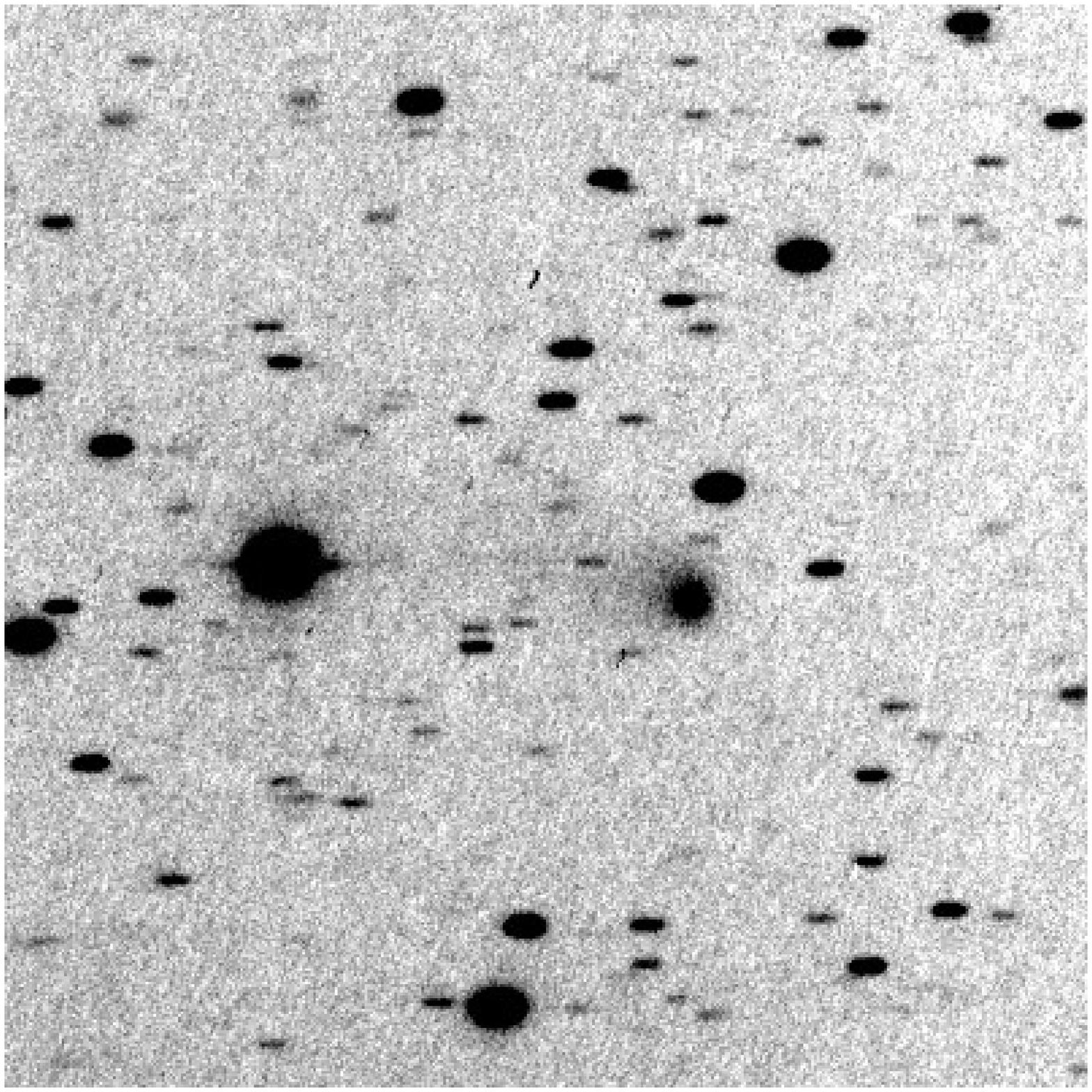}&
\includegraphics[height=4.64cm]{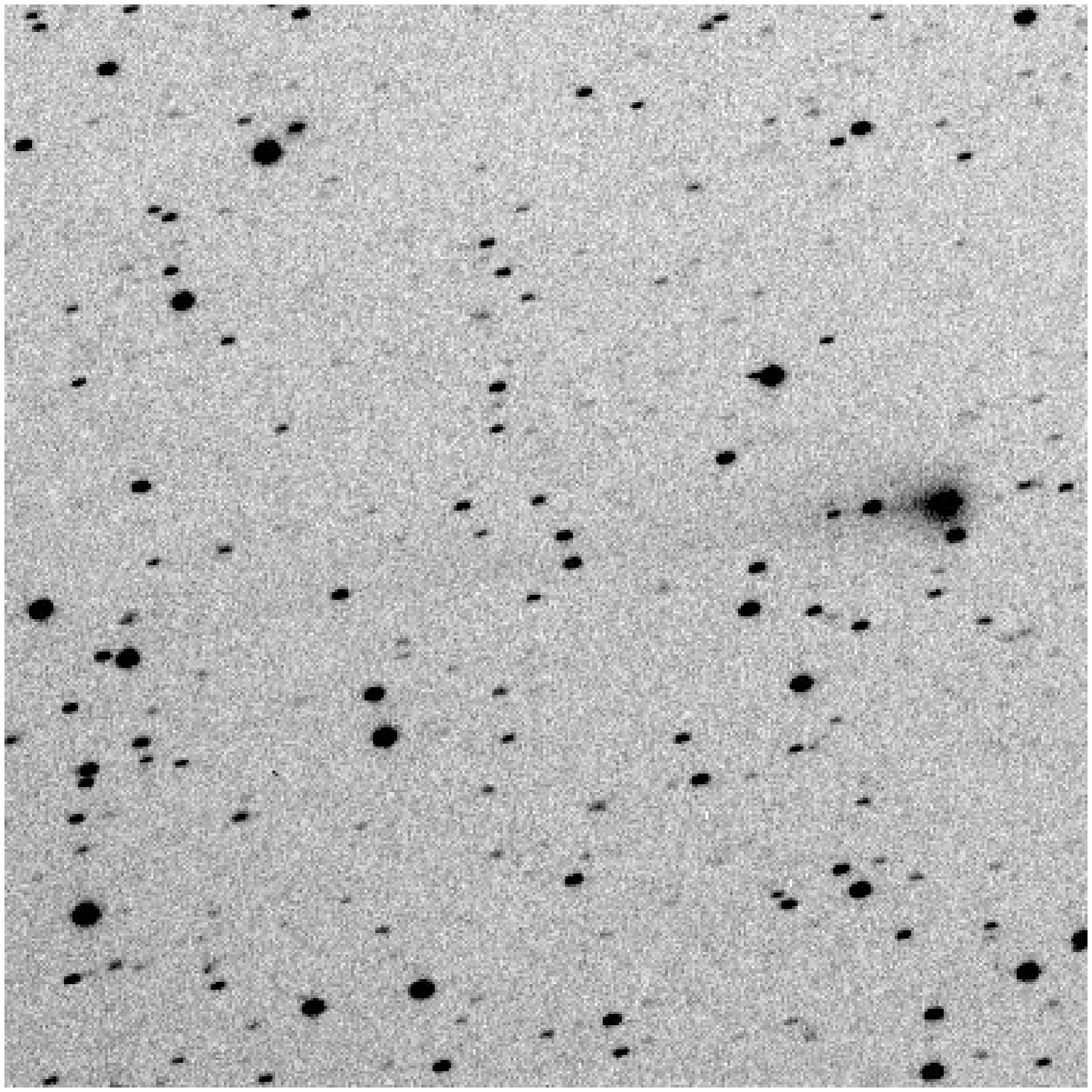}\\
\small C/2007 D1 [2008--12--29] & \small C/2007 D3 [2008--03--31] & \small C/2008 S3 [2008--12--31]\\
\includegraphics[height=4.64cm]{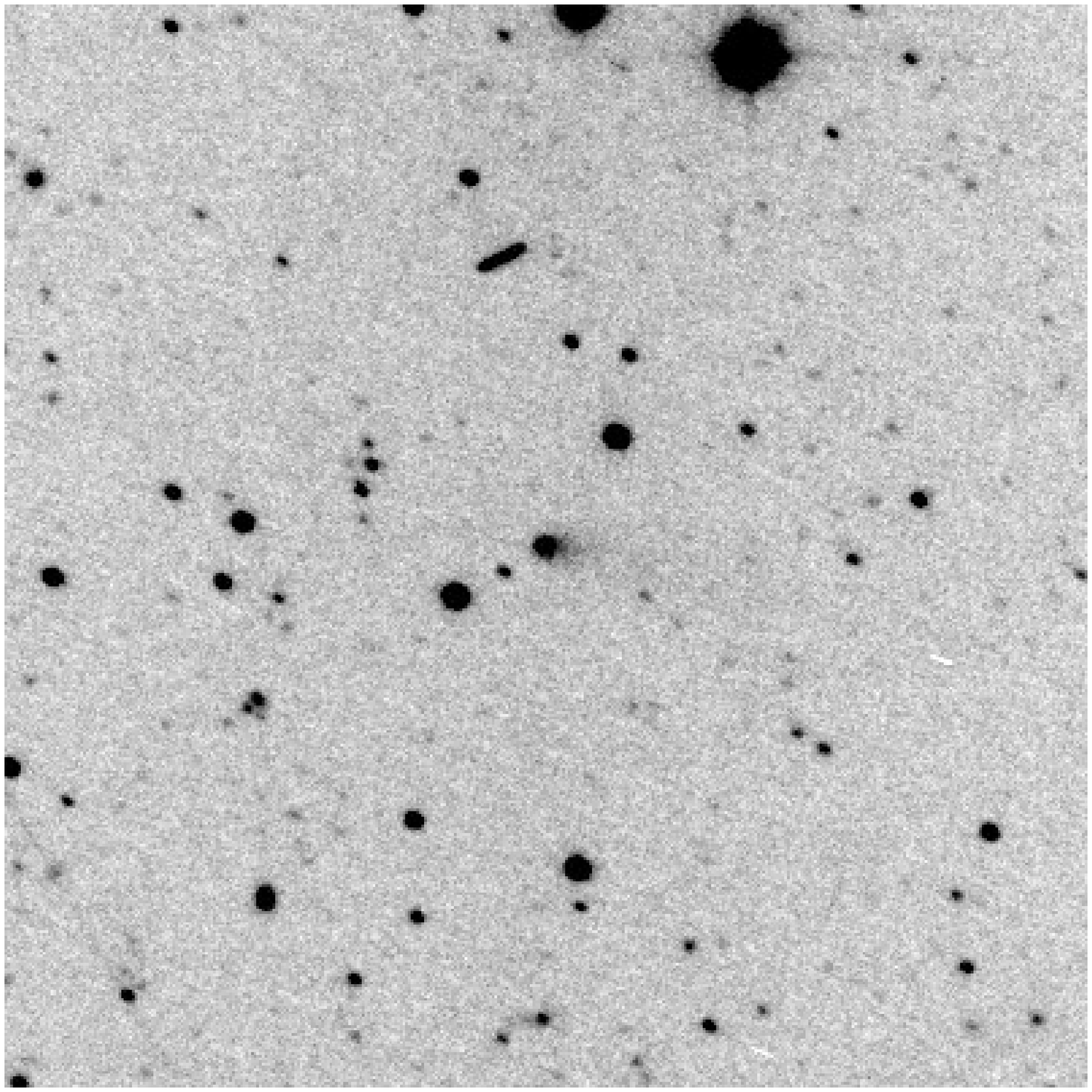}&
\noindent\includegraphics[height=4.64cm]{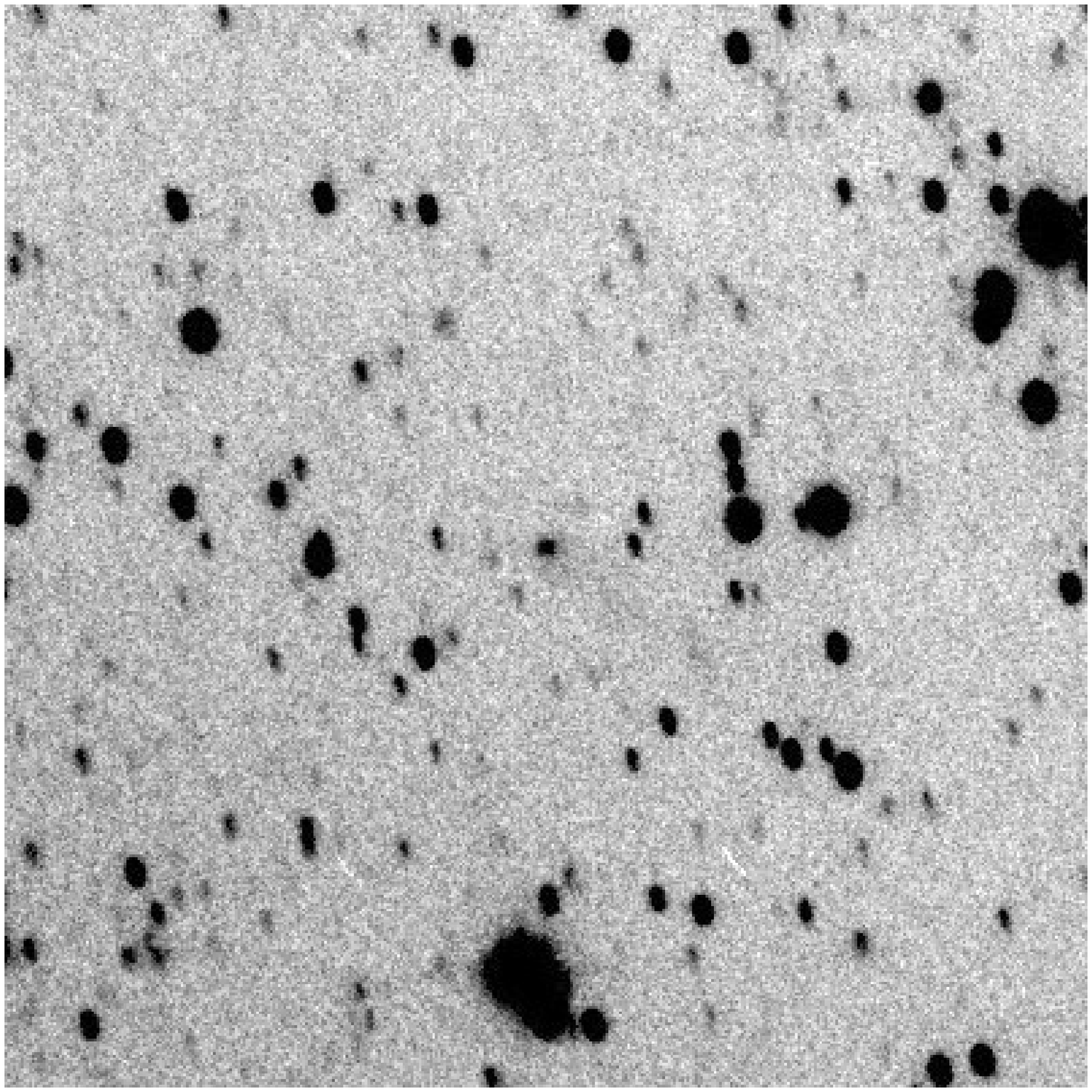}&
\includegraphics[height=4.64cm]{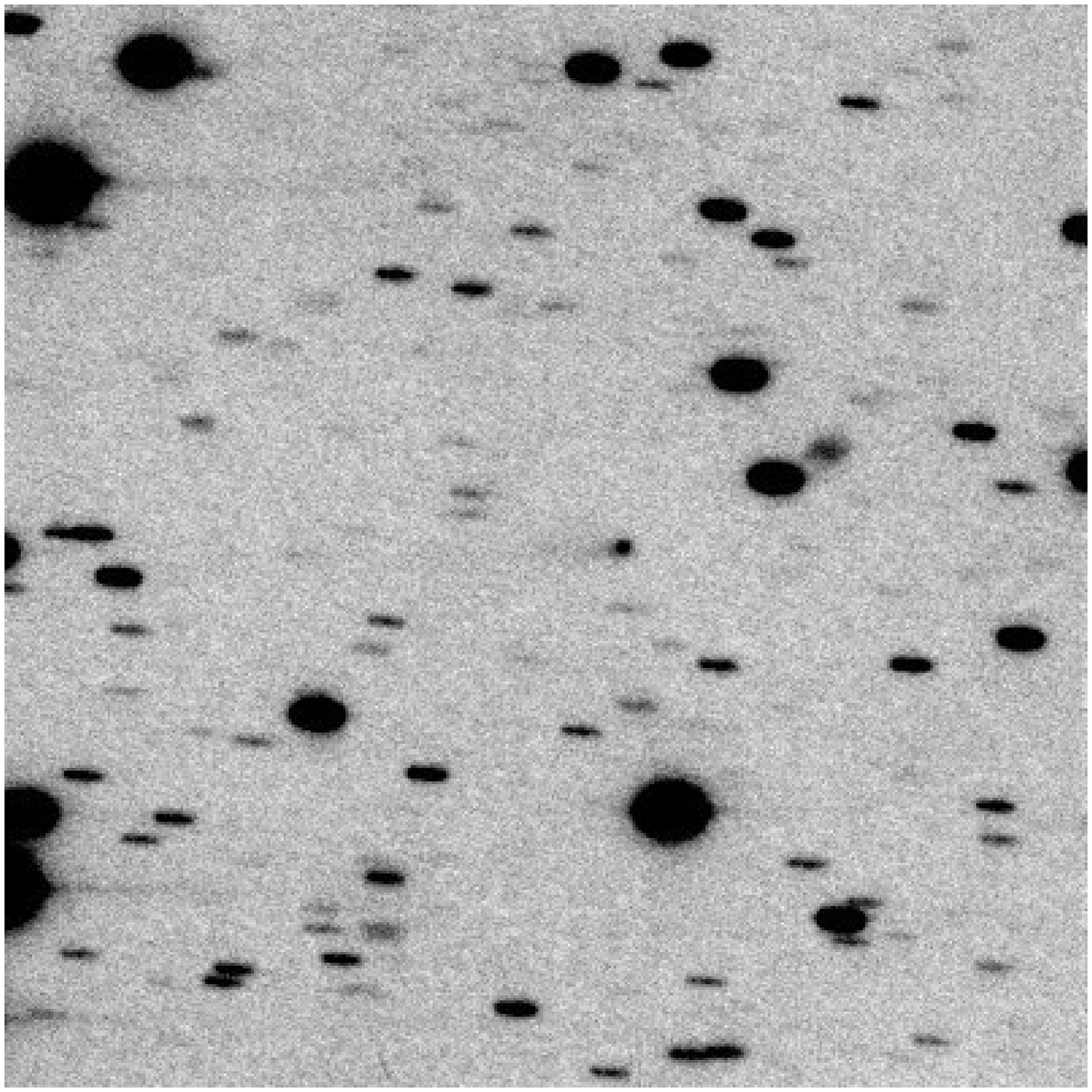}\\
\small C/2010 S1 [2013--07--13] & \small C/2012 K1 [2013--07--14] & \small C/2012 K8 [2014--07--25]\\
\includegraphics[height=4.64cm]{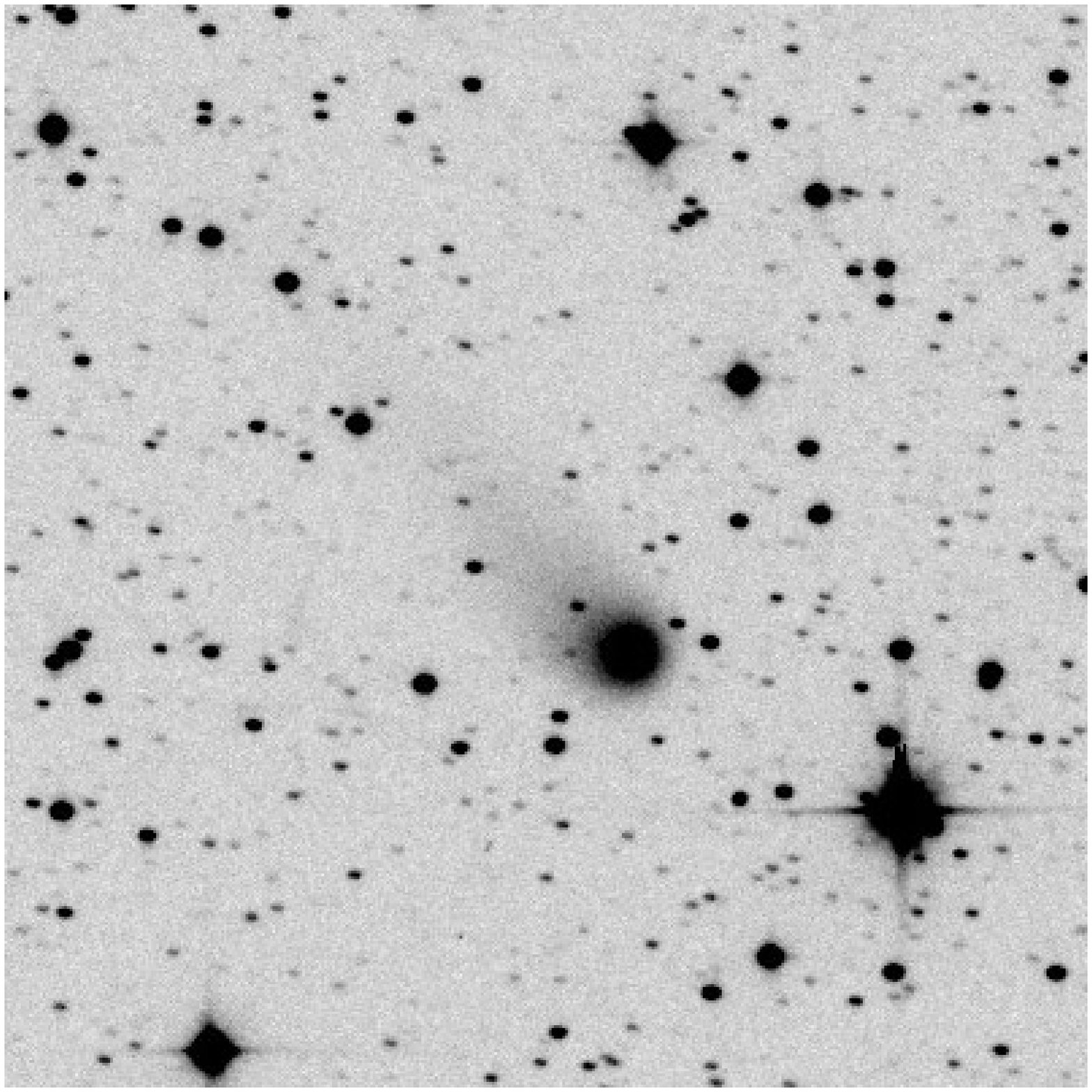}&
\noindent\includegraphics[height=4.64cm]{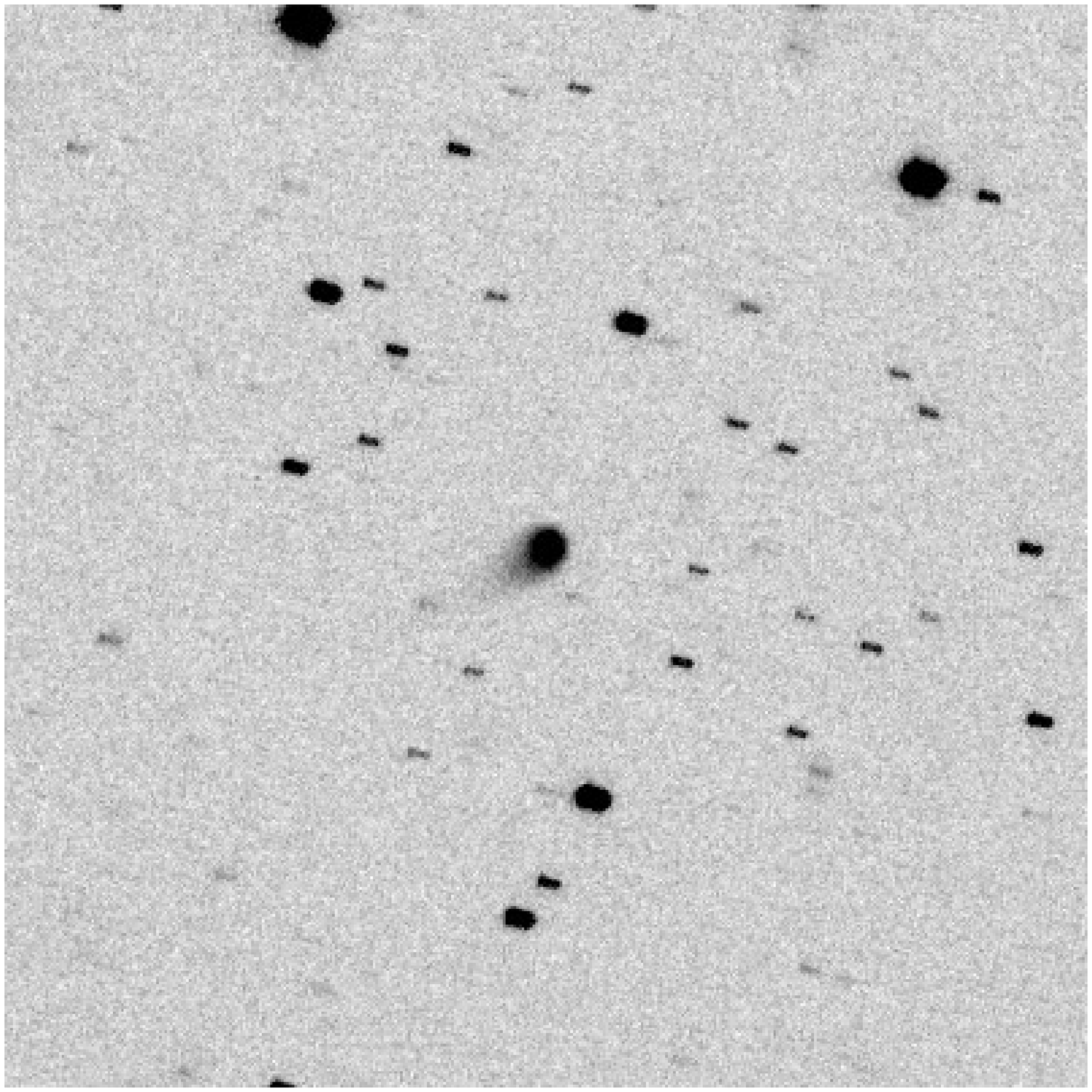}&
\includegraphics[height=4.64cm]{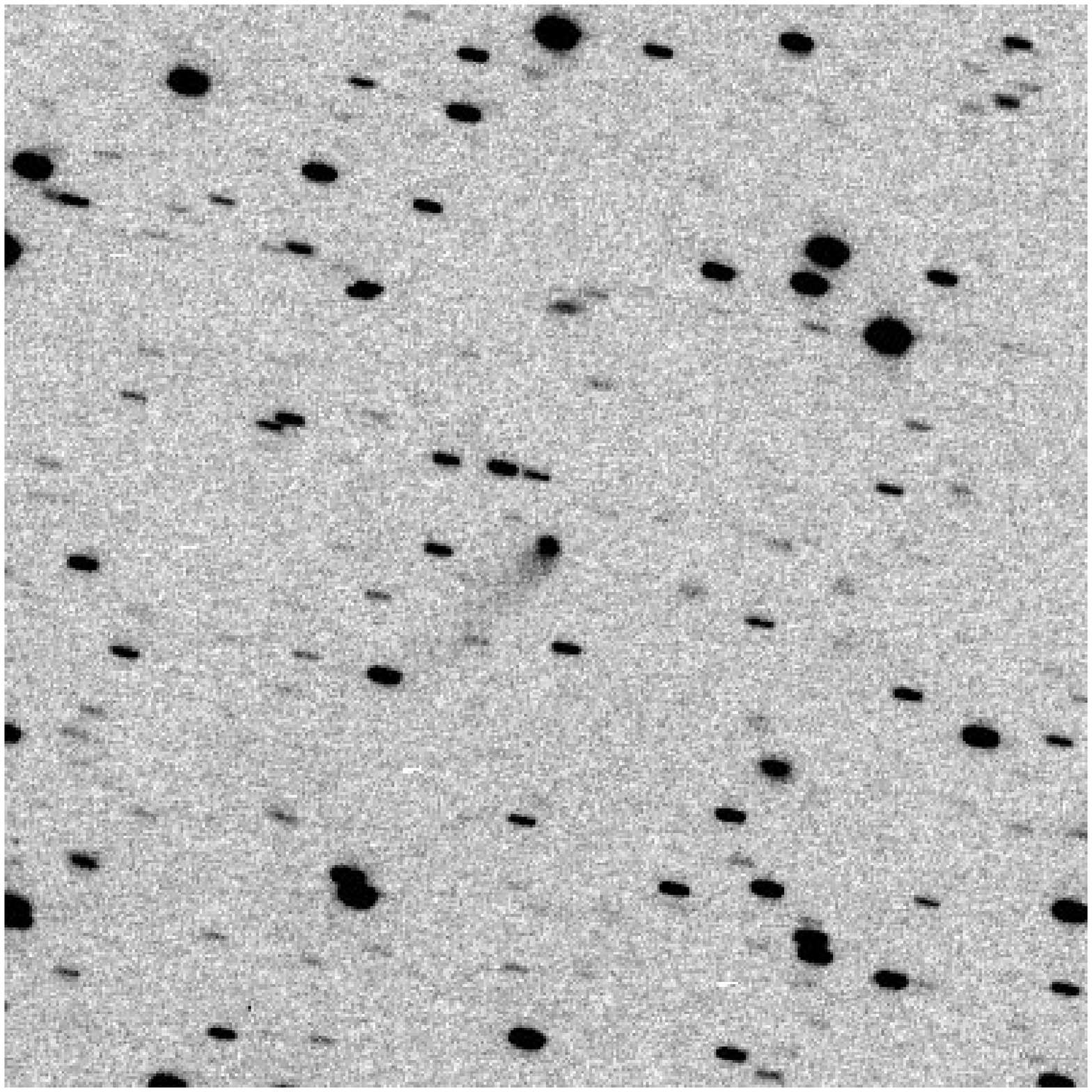}\\
\end{tabular}
\caption{An imagery of comets with dust tails. The top two rows show comets with extended tails, while the bottom two rows show comets with shorter and fainter tails, and these images have been centered to the comet. All fields of view are $6.7^\prime\times6.7^\prime$ for all comets but C/2005~L3 ($10^\prime\times10^\prime$).}
\end{figure*}

We carried out CCD observations on 103 nights from November, 1998 to October, 2014 at 
the Piszk\'estet\H{o}{} Station of the Konkoly Observatory, Hungary. Standard Johnson R filtered data were obtained using the 60/90/180 cm Schmidt-telescope. The list of observed comets is shown in Table 1, while the full log of observations is given in Table 2. Between November, 1998 and June, 2010 the telescope was equipped with a 
Photometrics  AT-200 CCD camera (1536x1024 pixel, KAF1600 MCII coated CCD chip). 
The projected area was $29^\prime\times18^\prime$, which corresponded to an angular resolution of 1.01$^{\prime\prime}$/pixel. 
Since 2010 August, an Apogee ALTA-U 4096$\times$4096 CCD camera (FoV $70^\prime\times70^\prime$ arcmin) with the same resolution. The operational temperature of the cameras was $-$40~$^\circ$C.

The exposure time was limited by the trailing of the comets during the
exposures. We restricted the maximal tolerable trails at $<2^{\prime\prime}$, 
which is characteristically the FWHM of the stellar profiles on an average
night.

During the first ten years, we observed distant comets occasionally, as a complement to the astrometric program of the Konkoly Observatory. Based on the initial results, in 2008 we started a dedicated program to observe distant LP comets at heliocentric distances larger than 5.2 AU, i.e. beyond the orbital distance of Jupiter.

The images were corrected in the standard fashion, stacked and evaluated with the Astrometrica software by H. Raab\footnote{astrometrica.at}. The photometric zero points were determined for the stacked images following our usual method: we first stacked the observed images to non-moving objects (``star'' images), measured the zero point of stellar photometry in this image, and then applied the same zero point to the ``comet'' images which were stacked taken the apparent motion into account. Stellar magnitudes were taken from the USNO B1 catalogue.

Since all comets were observed in activity and all had an apparently extended coma, we could determine the $Af\rho$ quantity and its slope (A'Hearn at al. 1984). 
This quantity measures the relative linear filling factor of dust and allows a comparison of data obtained at different sites, epochs, 
geometrical circumstances, and/or with different telescopes and photometric apertures. It is the product of Bond-albedo $A$
(Bond 1861; Bell 1917), the $f$ filling factor of the grains within the aperture and $\rho$ as the radius of the field of view at the comet, as
\begin{equation}
Af\rho = {({2DR})^2 \over \rho} \cdot {F_{com}\over F_{sol}}.
\end{equation}
where 
the Earth-comet distance $D$ and $\rho$ is in cm, the Sun-Earth distance $R$ is in AU, $F_{sol}$ is the flux of the Sun at 
$1$ AU referring to the photometric band used, $F_{com}$ is the observed flux from the comet.

$Af\rho$ is a function of $\rho$, usually slightly increasing in the coma and reaching the maximum value of $Af\rho_{max}$ at the
$\rho_{max}$ distance from the nucleus. Then it starts decreasing with a slope characteristic for the profile of the outer coma.
The average profile of the coma brightness is well described with a power-law as $rho^\gamma$, where gamma is the slope calculated by the $\log \rho$ dependence of $\log Af\rho$. Denoting the various coefficients simply with $K$ and $K'$, 
$\log Af\rho = K + \log \left( {1\over\rho} \int \rho\rho^\gamma {d}\rho \right) = K' + (\gamma+1)\log\rho,$ consequently 

\begin{equation}
slope = {d \log Af\rho \over  d \log \rho} = \gamma+1.
\end{equation}
 

Hence, the slope parameter has a value around $0$ if the coma is close to the steady state and globular symmetry (and $gamma=-1$ in this case). Deviations from this shape result in a non-zero slope parameter, either to the positive or the negative side. At larger heliocentric distances, comets are known to exhibit negative slope parameters, even down to $-1$, but there were also indications for temporal values of positive slope parameters (e.g. in the late activity of Hale-Bopp, Szab\'o et al. 2008).

The $Af\rho$ and slope parameters were determined from surface photometry profiles. The multiple aperture photometry (indicatively between 10,000--100,000 km) was plotted for all stacked images (in a log-log plot), and we manually selected the linear segment. We fitted a line in the log-log plane,  interpolated the \afrho{} to a 50,000 km aperture following the recipe of Milani et al. (2009), and also registered the slope parameter.  The peak of the profile ($\rho_{max}$) is often derived and handled as an independent morphological parameter. For sake of a direct comparison to previous studies we also determined $\rho_{max}$ and the corrseponding \afrhomax, i.e. \afrho at $\rho_{max}$. Nevertheless we note that we found a  strong correlation between the width of the stellar profiles and the $\rho_{max}$ values, showing that the smearing of the inner coma by the seeing biases the estimate of $\rho_{max}$, hence biasing \afrhomax as well.

\section{Results}
\setcounter{table}{2}
\begin{table} 
\caption{Summary of the results.}
\begin{tabular}{lcrrrrrc}
~~~~~~~~~~~~~~ Date~~~~~  &  R  [AU]    \afrho  [cm]   ~  Slope ~ I/O \cr
\hline
%
\end{tabular}
\begin{tabular}{l}
C/1997 BA6 (Spacewatch) \cr
\end{tabular}
\begin{tabular}{lcrrrrrc}
~~~~~~~~~~~~~~~~ 2003.09.19.  &  11.267 &   1047  &   0.16 & O \cr
\end{tabular}
\begin{tabular}{l}
C/1997 J1 (Mueller)     \cr 
\end{tabular}
\begin{tabular}{lcrrrrrc}
~~~~~~~~~~~~~~~~   1998.11.25.  &   6.061 &   83   &    0.14 & O \cr
\end{tabular}
\begin{tabular}{l}
C/1998 W3 (LINEAR)\cr
\end{tabular}
\begin{tabular}{lcrrrrrc}
~~~~~~~~~~~~~~~~    2000.03.10.  &   6.304 &   246   &  $-$0.01 & O \cr
\end{tabular}
\begin{tabular}{l}
C/1999 J2 (Skiff)\cr        
\end{tabular}
\begin{tabular}{lcrrrrrc}
~~~~~~~~~~~~~~~~   1999.09.24.  &   7.219 &   4931  &  $-$0.69 & I \cr
~~~~~~~~~~~~~~~~ 2000.01.01.  &   7.137 &   4090  &  $-$0.28 & I \cr
\end{tabular}
\begin{tabular}{l}
C/2002 V2 (LINEAR)\cr
\end{tabular}
\begin{tabular}{lcrrrrrc}
~~~~~~~~~~~~~~~~   2003.11.05.  &   6.909 &   453  &   $-$0.48 & O \cr
~~~~~~~~~~~~~~~~ 2003.11.09.  &   6.913 &   483  &   $-$0.42 & O \cr
\end{tabular}
\begin{tabular}{l}
C/2002 VQ94 (LINEAR)\cr
\end{tabular}
\begin{tabular}{lcrrrrrc}
~~~~~~~~~~~~~~~~   2008.03.31.  &   8.421 &   2113  &  $-$0.18 & O \cr
~~~~~~~~~~~~~~~~ 2008.12.29.  &   9.482 &   1754  &  $-$0.34 & O \cr
~~~~~~~~~~~~~~~~ 2009.04.16.  &   9.936 &   1733  &  $-$0.02 & O \cr
~~~~~~~~~~~~~~~~ 2010.01.15.  &  11.130 &   1661  &   0.20 & O \cr
\end{tabular}
\begin{tabular}{l}
C/2003 A2 (Gleason)\cr
\end{tabular}
\begin{tabular}{lcrrrrrc}
~~~~~~~~~~~~~~~~   2003.12.27.  &  11.429 &   2744  &   0.15 & O \cr
\end{tabular}
\begin{tabular}{l}
C/2003 K4 (LINEAR)\cr
\end{tabular}
\begin{tabular}{lcrrrrrc}
~~~~~~~~~~~~~~~~   2003.07.06.  &   5.762 &   1894  &  $-$0.95 & I \cr
~~~~~~~~~~~~~~~~ 2003.09.09.  &   5.145 &   1961  &  $-$0.18 & I \cr
\end{tabular}
\begin{tabular}{l}
C/2003 WT42 (LINEAR)\cr
\end{tabular}
\begin{tabular}{lcrrrrrc}
~~~~~~~~~~~~~~~~   2008.03.31.  &   7.383 &   271  &   $-$0.04 & O \cr
\end{tabular}
\begin{tabular}{l}
C/2004 B1 (LINEAR)\cr
\end{tabular}
\begin{tabular}{lcrrrrrc}
~~~~~~~~~~~~~~~~   2008.08.22.  &   9.134 &   232  &   $-$0.10 & O \cr
~~~~~~~~~~~~~~~~ 2008.12.30.  &  10.063 &   149  &   $-$0.13 & O \cr
\end{tabular}
\begin{tabular}{l}
C/2004 D1 (NEAT)\cr
\end{tabular}
\begin{tabular}{lcrrrrrc}
~~~~~~~~~~~~~~~~   2008.03.31.  &   7.600 &   131  &    0.15 & O \cr
\end{tabular}
\begin{tabular}{l}
C/2004 P1 (NEAT)\cr
\end{tabular}
\begin{tabular}{lcrrrrrc}
~~~~~~~~~~~~~~~~   2005.08.29.  &   7.909 &   196  &    0.06 & O \cr
~~~~~~~~~~~~~~~~ 2005.08.30.  &   7.913 &   224  &    0.07 & O \cr
\end{tabular}
\begin{tabular}{l}
C/2005 EL173 (LONEOS)\cr
\end{tabular}
\begin{tabular}{lcrrrrrc}
~~~~~~~~~~~~~~~~   2005.03.31.  &   6.944 &   23  &     0.08 & I \cr
~~~~~~~~~~~~~~~~ 2008.12.29.  &   6.705 &   449  &   $-$0.18 & O \cr
\end{tabular}
\begin{tabular}{l}
C/2005 G1 (LINEAR)\cr
\end{tabular}
\begin{tabular}{lcrrrrrc}
~~~~~~~~~~~~~~~~   2005.04.03.  &   5.566 &   1473  &  $-$0.42 & I \cr
\end{tabular}
\begin{tabular}{l}
C/2005 L3 (McNaught)\cr
\end{tabular}
\begin{tabular}{lcrrrrrc}
~~~~~~~~~~~~~~~~   2008.05.28.  &   5.677 &   8966  &  $-$0.74 & O \cr
~~~~~~~~~~~~~~~~ 2008.12.31.  &   6.136 &   6122  &  $-$0.44 & O \cr
~~~~~~~~~~~~~~~~ 2009.04.18.  &   6.486 &   6480  &  $-$0.61 & O \cr
~~~~~~~~~~~~~~~~ 2009.05.01.  &   6.535 &   5779  &  $-$0.42 & O \cr
~~~~~~~~~~~~~~~~ 2009.12.29.  &   7.525 &   4774  &  $-$0.23 & O \cr
~~~~~~~~~~~~~~~~ 2010.02.22.  &   7.776 &   4133  &  $-$0.16 & O \cr
~~~~~~~~~~~~~~~~ 2010.05.23.  &   8.203 &   3395  &   0.02 & O \cr
~~~~~~~~~~~~~~~~ 2010.06.12.  &   8.299 &   3325  &  $-$0.29 & O \cr
~~~~~~~~~~~~~~~~ 2014.02.24.  &  15.152 &   1452  &   0.55 & O \cr
\end{tabular}
\end{table}
\setcounter{table}{2}
\begin{table}
\caption{cont.}
\begin{tabular}{lcrrrrrc}
~~~~~~~~~~~~~~ Date~~~~~  &  R  [AU]    \afrho  [cm]   ~  Slope ~ I/O \cr
\hline
%
\end{tabular}
\begin{tabular}{l}
C/2005 S4 (McNaught)\cr
\end{tabular}
\begin{tabular}{lcrrrrrc}
~~~~~~~~~~~~~~~~   2008.05.26.  &   6.252 &   215  &    0.07 & O \cr
~~~~~~~~~~~~~~~~ 2008.08.20.  &   6.490 & 200  &   $-$0.08 & O \cr
~~~~~~~~~~~~~~~~ 2008.08.26.  &   6.508 &   217  &   $-$0.18 & O \cr
~~~~~~~~~~~~~~~~ 2009.04.02.  &   7.282 &   148  &    0.03 & O \cr
~~~~~~~~~~~~~~~~ 2009.05.03.  &   7.411 &   217  &    0.07 & O \cr
\end{tabular}
\begin{tabular}{l}
C/2006 A2 (Catalina)\cr
\end{tabular}
\begin{tabular}{lcrrrrrc}
~~~~~~~~~~~~~~~~   2006.01.22.  &   5.625 &   95  &    $-$0.53 & O \cr
\end{tabular}
\begin{tabular}{l}
C/2006 K1 (McNaught)\cr
\end{tabular}
\begin{tabular}{lcrrrrrc}
~~~~~~~~~~~~~~~~   2008.10.25.  &   5.762 &   332  &   $-$0.18 & O \cr
~~~~~~~~~~~~~~~~ 2008.11.28.  &   5.931 &   252  &   $-$0.30 & O \cr
~~~~~~~~~~~~~~~~ 2008.11.28.  &   5.935 &   280  &   $-$0.15 & O \cr
~~~~~~~~~~~~~~~~ 2008.12.28.  &   6.088 &   279  &   $-$0.54 & O \cr
\end{tabular}
\begin{tabular}{l}
C/2006 K4 (NEAT)\cr
 \end{tabular}
\begin{tabular}{lcrrrrrc}
~~~~~~~~~~~~~~~~  2006.06.20.  &   5.708 &   146  &   $-$0.07 & I \cr
\end{tabular}
\begin{tabular}{l}
C/2006 M2 (Spacewatch)\cr
\end{tabular}
\begin{tabular}{lcrrrrrc}
~~~~~~~~~~~~~~~~   2006.06.20.  &   5.445 &   36  &    $-$0.88 & O \cr
\end{tabular}
\begin{tabular}{l}
C/2006 S3 (LONEOS)\cr
\end{tabular}
\begin{tabular}{lcrrrrrc}
~~~~~~~~~~~~~~~~   2006.09.23.  &  14.265 &   1926  &  $-$0.13 & I \cr
~~~~~~~~~~~~~~~~ 2008.08.25.  &  10.563 &   1530  &  $-$0.29 & I \cr
~~~~~~~~~~~~~~~~ 2008.12.28.  &   9.894 &   2468  &  $-$0.15 & I \cr
~~~~~~~~~~~~~~~~ 2009.12.26.  &   7.975 &   2178  &  $-$0.15 & I \cr
~~~~~~~~~~~~~~~~ 2010.06.12.  &   7.137 &   2380  &  $-$0.51 & I \cr
~~~~~~~~~~~~~~~~ 2014.02.23.  &   7.16  &   7003  &  $-$0.17 & I \cr
\end{tabular}
\begin{tabular}{l}
C/2007 B2 (Skiff)\cr
 \end{tabular}
\begin{tabular}{lcrrrrrc}
~~~~~~~~~~~~~~~~  2007.03.06.  &   5.733 &   332  &   $-$0.26 & I \cr
\end{tabular}
\begin{tabular}{l}
C/2007 D1 (LINEAR)\cr
\end{tabular}
\begin{tabular}{lcrrrrrc}
~~~~~~~~~~~~~~~~   2008.12.30.  &   9.369 &   1054  &  $-$0.26 & O \cr
~~~~~~~~~~~~~~~~ 2009.04.16.  &   9.597 &   1649  &  $-$0.37 & O \cr
~~~~~~~~~~~~~~~~ 2009.12.29.  &  10.259 &   1167  &  $-$0.22 & O \cr
\end{tabular}
\begin{tabular}{l}
C/2007 D3 (LINEAR)\cr
\end{tabular}
\begin{tabular}{lcrrrrrc}
~~~~~~~~~~~~~~~~   2008.03.31.  &   5.695 &   227  &    0.003 & O \cr
~~~~~~~~~~~~~~~~ 2008.12.24.  &   6.706 &   138  &   $-$0.27 & O \cr
~~~~~~~~~~~~~~~~ 2009.04.18.  &   7.239 &   164  &   $-$0.02 & O \cr
\end{tabular}
\begin{tabular}{l}
C/2007 G1 (LINEAR)\cr
\end{tabular}
\begin{tabular}{lcrrrrrc}
~~~~~~~~~~~~~~~~   2007.05.03.  &   5.979 &   629  &   $-$0.43 & I \cr
\end{tabular}
\begin{tabular}{l}
C/2007 JA21 (LINEAR)\cr
\end{tabular}
\begin{tabular}{lcrrrrrc}
~~~~~~~~~~~~~~~~  2008.04.01.  &   6.506 &   184  &   $-$0.34 & O \cr
~~~~~~~~~~~~~~~~ 2008.08.27.  &   7.143 &   219  &   $-$0.40 & O \cr
\end{tabular}
\begin{tabular}{l}
C/2007 K1 (Lemmon)\cr
\end{tabular}
\begin{tabular}{lcrrrrrc}
~~~~~~~~~~~~~~~~   2008.05.28.  &   9.487 &   243  &   $-$0.51 & O \cr
~~~~~~~~~~~~~~~~ 2008.08.20.  &   9.605 &   381  &    0.01 & O \cr
~~~~~~~~~~~~~~~~ 2008.08.24.  &   9.611 &   389  &    0.07 & O \cr
~~~~~~~~~~~~~~~~ 2009.04.20.  &  10.052 &   191  &   $-$0.86 & O \cr
\end{tabular}
\begin{tabular}{l}
C/2007 M1 (McNaught)\cr
 \end{tabular}
\begin{tabular}{lcrrrrrc}
~~~~~~~~~~~~~~~~  2008.08.24.  &   7.475 &   486  &    0.12 & O \cr
~~~~~~~~~~~~~~~~ 2009.05.04.  &   7.657 &   480  &   $-$0.00 & O \cr
~~~~~~~~~~~~~~~~ 2010.06.12.  &   8.551 &   316  &   $-$0.87 & O \cr
\end{tabular}
\end{table}
\setcounter{table}{2}
\begin{table}
\caption{cont.}
\begin{tabular}{l}
C/2007 U1 (LINEAR)\cr
\end{tabular}
\begin{tabular}{lcrrrrrc}
~~~~~~~~~~~~~~~~   2010.06.13.  &   6.721 &   115  &   $-$0.33 & O \cr
\end{tabular}
\begin{tabular}{l}
C/2007 VO53 (Spacewatch)\cr
\end{tabular}
\begin{tabular}{lcrrrrrc}
~~~~~~~~~~~~~~~~   2008.08.21.  &   6.707 &   136  &   $-$0.28 & I \cr
~~~~~~~~~~~~~~~~ 2008.10.26.  &   6.387 &   161  &    0.03 & I \cr
~~~~~~~~~~~~~~~~ 2008.12.23.  &   6.115 &   193  &   $-$0.25 & I \cr
~~~~~~~~~~~~~~~~ 2009.09.16.  &   5.142 &   593  &   $-$0.41 & I \cr
\end{tabular}
\begin{tabular}{l}
C/2008 S3 (Boattini)\cr
\end{tabular}
\begin{tabular}{lcrrrrrc}
~~~~~~~~~~~~~~~~   2008.10.25.  &   9.829 &   554  &    0.12 & I \cr
~~~~~~~~~~~~~~~~ 2008.12.30.  &   9.611 &   784  &   $-$0.37 & I \cr
~~~~~~~~~~~~~~~~ 2009.09.15.  &   8.867 &   827  &   $-$0.76 & I \cr
~~~~~~~~~~~~~~~~ 2013.09.27.  &   9.470 &   924  &   $-$0.20 & O \cr
~~~~~~~~~~~~~~~~ 2014.07.25.  &  10.493 &   646  &   $-$0.21 & O \cr
~~~~~~~~~~~~~~~~ 2014.09.29.  &  10.737 &   657  &   $-$0.28 & O \cr
\end{tabular}
\begin{tabular}{l}
C/2010 D4 (WISE)\cr
\end{tabular}
\begin{tabular}{lcrrrrrc}
~~~~~~~~~~~~~~~~ \end{tabular}
\begin{tabular}{lcrrrrrc}
~~~~~~~~~~~~~~~~   2010.03.24.  &   7.469 &   133  &   $-$0.63 & O \cr
\end{tabular}
\begin{tabular}{l}
C/2010 G2 (Hill)\cr
\end{tabular}
\begin{tabular}{lcrrrrrc}
~~~~~~~~~~~~~~~~   2010.05.01.  &   5.457 &   147  &   $-$0.18 & I \cr
\end{tabular}
\begin{tabular}{l}
C/2010 R1 (LINEAR)\cr
\end{tabular}
\begin{tabular}{lcrrrrrc}
~~~~~~~~~~~~~~~~   2012.03.18.  &   5.639 &   907  &   $-$0.48 & I \cr
~~~~~~~~~~~~~~~~ 2014.02.24.  &   7.253 &   267  &   $-$0.19 & O \cr
\end{tabular}
\begin{tabular}{l}
C/2010 S1 (LINEAR)\cr
\end{tabular}
\begin{tabular}{lcrrrrrc}
~~~~~~~~~~~~~~~~   2013.10.02.  &   5.977 &   9154  &  $-$0.45 & O \cr
~~~~~~~~~~~~~~~~ 2014.03.15.  &   6.265 &   9106  &  $-$0.35 & O \cr
~~~~~~~~~~~~~~~~ 2014.10.21.  &   6.927 &   8740  &  $-$0.12 & O \cr
\end{tabular}
\begin{tabular}{l}
C/2010 U3 (Boattini)\cr
 \end{tabular}
\begin{tabular}{lcrrrrrc}
~~~~~~~~~~~~~~~~  2013.12.27.  &  13.563 &   2539  &   0.27 & I \cr
~~~~~~~~~~~~~~~~ 2014.07.27.  &  12.715 &   1353  &   0.15 & I \cr
~~~~~~~~~~~~~~~~ 2014.10.20.  &  12.381 &   1729  &   1.17 & I \cr
\end{tabular}
\begin{tabular}{l}
C/2011 F1 (LINEAR)\cr
\end{tabular}
\begin{tabular}{lcrrrrrc}
~~~~~~~~~~~~~~~~   2011.03.25.  &   6.926 &   957  &   $-$0.42 & I \cr
\end{tabular}
\begin{tabular}{l}
C/2011 KP36 (Spacewatch)\cr
\end{tabular}
\begin{tabular}{lcrrrrrc}
~~~~~~~~~~~~~~~~   2013.08.05.  &   8.478 &   1435 &   $-$0.19 & I \cr
~~~~~~~~~~~~~~~~ 2013.09.28.  &   8.212 &   1141 &    0.03 & I \cr
\end{tabular}
\begin{tabular}{l}
C/2012 K1 (PANSTARRS)\cr
\end{tabular}
\begin{tabular}{lcrrrrrc}
~~~~~~~~~~~~~~~~   2012.05.20.  &   8.784 &   1216  &  $-$0.35 & I \cr
~~~~~~~~~~~~~~~~ 2013.03.12.  &   6.352 &   2466  &  $-$0.91 & I \cr
~~~~~~~~~~~~~~~~ 2013.07.14.  &   5.210 &   2738  &  $-$0.65 & I \cr
\end{tabular}
\begin{tabular}{l}
C/2012 K8 (Lemmon)\cr
\end{tabular}
\begin{tabular}{lcrrrrrc}
~~~~~~~~~~~~~~~~   2013.09.27.  &   6.826 &   574  &   $-$0.67 & I \cr
~~~~~~~~~~~~~~~~ 2014.07.25.  &   6.465 &   769  &   $-$0.38 & I \cr
\end{tabular}
\begin{tabular}{l}
C/2012 LP26 (Palomar)\cr
\end{tabular}
\begin{tabular}{lcrrrrrc}
~~~~~~~~~~~~~~~~   2013.08.01.  &   8.172 &   340  &   $-$0.78 & I \cr
~~~~~~~~~~~~~~~~ 2014.06.03.  &   7.162 &   579  &   $-$0.05 & I \cr
~~~~~~~~~~~~~~~~ 2014.07.26.  &   7.026 &   615  &   $-$0.58 & I \cr
~~~~~~~~~~~~~~~~ 2014.10.21.  &   6.836 &   568  &   $-$0.42 & I \cr
\end{tabular}
\end{table}
\setcounter{table}{2}
\begin{table}
\caption{cont.}
\begin{tabular}{l}
C/2012 S1 (ISON)\cr
\end{tabular}
\begin{tabular}{lcrrrrrc}
~~~~~~~~~~~~~~~~   2012.10.21.  &   5.997 &   760  &   $-$0.43 & I \cr
~~~~~~~~~~~~~~~~ 2012.11.14.  &   5.639 &   948  &   $-$0.66 & I \cr
\end{tabular}
\begin{tabular}{l}
C/2012 U1 (PANSTARRS)\cr
\end{tabular}
\begin{tabular}{lcrrrrrc}
~~~~~~~~~~~~~~~~   2013.01.07.  &   6.601 &   199  &   $-$0.09 & I \cr
~~~~~~~~~~~~~~~~ 2013.07.12.  &   5.895 &   253  &    0.28 & I \cr
\end{tabular}
\begin{tabular}{l}
C/2013 G9 (Tenagra)\cr
\end{tabular}
\begin{tabular}{lcrrrrrc}
~~~~~~~~~~~~~~~~   2013.06.09.  &   6.822 &   394  &   $-$0.79 & I \cr
~~~~~~~~~~~~~~~~ 2013.07.02.  &   6.723 &   467  &   $-$0.29 & I \cr
~~~~~~~~~~~~~~~~ 2014.03.14.  &   5.799 &   938  &   $-$0.67 & I \cr
\end{tabular}
\begin{tabular}{l}
C/2013 L2 (Catalina)\cr
\end{tabular}
\begin{tabular}{lcrrrrrc}
~~~~~~~~~~~~~~~~   2013.06.08.  &   5.731 &   173  &   $-$0.54 & O \cr
~~~~~~~~~~~~~~~~ 2013.06.12.  &   5.750 &   170  &   $-$0.05 & O \cr
~~~~~~~~~~~~~~~~ 2013.07.11.  &   5.868 &   176  &   $-$0.24 & O \cr
~~~~~~~~~~~~~~~~ 2013.08.12.  &   6.003 &   173  &   $-$0.51 & O \cr
\end{tabular}
\begin{tabular}{l}
C/2013 P3 (Palomar)\cr
\end{tabular}
\begin{tabular}{lcrrrrrc}
~~~~~~~~~~~~~~~~   2013.08.16.  &   9.059 &   620  &   $-$0.25 & I \cr
~~~~~~~~~~~~~~~~ 2013.09.27.  &   8.990 &   905  &   $-$0.43 & I \cr
~~~~~~~~~~~~~~~~ 2013.10.31.  &   8.938 &   1012  &  $-$0.56 & I \cr
~~~~~~~~~~~~~~~~ 2013.12.04.  &   8.890 &   1097  &  $-$0.63 & I \cr
~~~~~~~~~~~~~~~~ 2014.07.25.  &   8.676 &   1071  &  $-$0.79 & I \cr
~~~~~~~~~~~~~~~~ 2014.10.25.  &   8.648 &   1438  &  $-$0.64 & I \cr
\end{tabular}
\begin{tabular}{l}
C/2013 X1 (PANSTARRS)\cr
\end{tabular}
\begin{tabular}{lcrrrrrc}
~~~~~~~~~~~~~~~~   2013.12.27.  &   8.712 &   451  &   $-$0.79 & I \cr
~~~~~~~~~~~~~~~~ 2014.03.14.  &   8.119 &   557  &   $-$0.20 & I \cr
~~~~~~~~~~~~~~~~ 2014.08.23.  &   6.817 &   1067  &  $-$0.79 & I \cr
~~~~~~~~~~~~~~~~ 2014.08.25.  &   6.801 &   1153  &  $-$0.81 & I \cr
~~~~~~~~~~~~~~~~ 2014.10.28.  &   6.255 &   1400  &  $-$0.69 & I \cr
\end{tabular}
\begin{tabular}{l}
C/2014 L5 (Lemmon)\cr
\end{tabular}
\begin{tabular}{lcrrrrrc}
~~~~~~~~~~~~~~~~   2014.10.27.  &   6.207 &   127  &   $-$0.63 & I \cr
\end{tabular}
\begin{tabular}{l}
C/2014 M2 (Christensen)\cr
\end{tabular}
\begin{tabular}{lcrrrrrc}
~~~~~~~~~~~~~~~~ \end{tabular}
\begin{tabular}{lcrrrrrc}
~~~~~~~~~~~~~~~~   2014.09.29.  &   6.925 &   084  &   $-$0.70 & O \cr
\end{tabular}
\begin{tabular}{l}
C/2014 R3 (PANSTARRS)\cr
\end{tabular}
\begin{tabular}{lcrrrrrc}
~~~~~~~~~~~~~~~~     2014.10.28.  &   8.344 &   459  &   $-$0.47 & I \cr
\hline                                                                                            
\end{tabular}                                                                                     
\end{table}

The summary of the reduced observations (\afrho{} and slope parameters) is given in Table 3. While most targets have no published history in the literature, some comets in the sample have extensive observational records. Here we give a detailed discussion about those comets which have published independent \afrho measurements, or our observations cover more than 1~AU in heliocentric distance at least. Morphological parameters were determined following the metodf we described in Szab\'o et al. 2001 and 2002.

\bigskip
\noindent
{\it C/2002 VQ94 (LINEAR)}
\newline
\noindent
This long period comet with an orbital period of 2875 years was discovered as
an asteroid by the LINEAR team on November 11, 2002 at a heliocentric distance of 10.0 AU. The diameter of the nucleus was estimated to be 80~km (Jewitt 2005) to 96~km (Korsun et al. 2014). Observations from the end of August 2003, when the heliocentric distance of the comet was 8.9 AU, revealed a prominent coma with a fan-like morphology, suggesting that large ($>$~10~$\mu$m) grains were not produced (Ivanova et al. 2011). The discovery was made almost three and a half years before the perihelion at 5.20 AU, when the initial brightness of 18\fm5 brightened up to 16\fm5. The comet was found to be rich in $CO^+$ near perihelium, while the measured $N^+/CO^+$ ratio was suggested to be an evidence for that this comet was formed in a cold environment (Korsun et al. 2006). The late activity was characterized by $Af\rho$ values of 2000 cm in 2008, and 800 cm in 2009 and 2011 (Korsun et al. 2014).

We observed this comet on 7 epochs, on the inward orbit at R=8.58 AU, and on the outward orbit between 8.30 and 11.13 AU. $Af\rho$ values varied between 1300 and 2200 cm, with only little dependence on the heliocentric distance and apparently more influenced by temporal changes. The slope was continuously slighty negative, with values between $-0.3$..$-0.1$. The appearance of the comet was generally similar to the detections reported in the literature, showing a bright photocenter and a fan-like elongated coma. However, in 2003 November and 2010 January, only the photocenter was detected without a coma  above the background scatter.

\bigskip
\noindent
{\it C/2005 L3 (McNaught)}
\newline
\noindent
The comet was discovered about two and a half years prior to perihelion (5.6 AU) by Robert McNaught in the Siding Spring Survey project in June 2005, at 8.7 AU heliocentric distance. Further prediscovery images were found from the previous year by McNaught, capturing the comet at 10.4 AU.  This is a dynamically new comet from the Oort cloud, according to the classification proposed by Levison (1996), because $a > 10^4$ AU. The derived absolute brightness is about H$_{10} = 3$\fm0 mag, which is quite bright compared to other similar comets. The perihelion passage occurred at 5.58 AU in 2008 January, at a maximum visual brightness of 13\fm5. Recently, Mazzotta Epifani et al. (2014) derived an $Af\rho_{max}=5255\pm 46$ cm at $R=6.64$~AU. They detected a
large twisted structure in the radial-normalized image, extending mostly to the south, and a well-defined jet-like structure in the Laplace-filtered image. They interpreted these findings as a possible hint for an active area and a nucleus rotation effect.

We observed this comet on the outward orbit, between 5.62 and 15.12 AU. We detected the decreasing trend of \afrho{} between 9450--1530~cm, in accordance with the previous measurements (e.g. we derived 6084 cm on 2009--05--01, on a decreasing trend with the increasing solar distance, while Mazzotta Epifani et al. (2014) published 5255~cm at 2009--05--29). A very prominent coma and tail were continuously observed for this comet during 2008--2010. The length of the tail was measured between 4--9$^\prime$. The longest tails were detected in 2009 December (9$^\prime$) and 2008 December (8$^\prime$). The appearance drastically changed by 2011, when only a very faint diffuse tail was detectable above the background scatter.

\bigskip
\noindent
{\it C/2005 S4 (McNaught)}
\newline
\noindent
Robert McNaught discovered this returning comet ($P\approx$125,000 years) in
Sept. 2005 at a heliocentric distance of 7.4 AU, about two years prior to perihelion passage ($q=5.9$~AU). The initial brightness of 18\fm5 brightened up to 16\fm0, while the absolute brightness was about H$_{10} = 5$\fm5 mag during the inward phase. Mazzotta Epifani et al. (2014) derived an $Af\rho_{max} = 116\pm 2$ cm at R = 7.52 AU, when the coma was surrounded by a faint, tail-like feature. 

We observed this comet on the outward orbit, between 6.25 and 7.41 AU. The appearance was generally similar to that of described by Mazzotta Epifani et al. (2014); the diffuse coma with a few arc seconds diameter was completed by a faint tail, measured between 30$^{\prime\prime}$ (2008--08--22) and 2$^\prime$ (2009--04--01). On 2009--05--03, the situation changed dramaticaly: the photocenter and the coma dimmed significantly, while the tail seemed to be relatively brighter, still compatible in brightness to the images from the previous weeks. The overall view suggested that the activity was in cessation at the time of our last observation, while the tail was still relatively rich in the matter ejected around the end of the activity.

\bigskip
\noindent
{\it C/2006 S3 (LONEOS)}
\newline
\noindent
This dynamically new comet was discovered at a brightness of 19\fm0 by the Lowell Observatory Near-Earth Object Search (LONEOS) project in Sept. 2006 at record distance (14.4 AU) and record early, about five and half years prior to 
perihelion. Prediscovery images were found on Catalina Sky Survey frames from the previous month. Further orbital calculations provided a perihelion distance of 5.13 AU and a perihelion passage in April 2012. The maximum brightness was around 11\fm0, the derived absolute brightness is about H$_{10} = 2$\fm0 mag, which is the second brightest in our sample. Recently, Shi et al. (2014) presented a few photometric data of this target, obtained pre-perihelion at 5.86 AU. 

Rousselot et al. (2014) performed both spectroscopic and photometric monitoring of the cometary activity during 8 years, between December 2006 and March 2014. They found a lack of emission lines, indicating a dusty comet, and suggested that $Af\rho$ values were larger in post-perihelion relative to the pre-perihelion measurements at similar heliocentric distances. They determined $Af\rho$ values in the range of 700--4000 cm, and considering the long-lasting activity, suggested that this comet represented one of the most dust-productive objects in the Solar System. Our measurements nicely complete the dataset of Rousselot et al. (2014). 

In 2006, we detected an elongated coma toward NW, with a dust filling factor of $Af\rho=513$~cm. In 2008--2009, we could detect a faint tail of $\approx 30$--$60^{\prime\prime}$ length, leaving the bright coma toward E, and NE in 2009. The $Af\rho$ values followed an increasing trend, and were compatible with those of determined by Rousselot et al. (2014). In 2010, the comet kept brightening, while the length of the tail evolved up to 2--3$^\prime$. In 2014, the appearance drastically changed, on February 24, the coma was unexpctedly bright, and we observed a 10$^\prime$ long dust tail toward PA~100$\pm 1^\circ$. On that day, we derived an extraordinary $Af\rho=7003$~cm, exceeding all the previous determinations, and being also a factor of 2 larger than that of derived by Rousselot et al. (2014) on 2014--03--02. The environment of the comet was clear of background stars in our image, therefore we expect that the increased $Af\rho$ value reflects the interesting behaviour of this exciting comet, rather than an observational artifact.

\begin{figure*}
\includegraphics[angle=270,width=8.0cm]{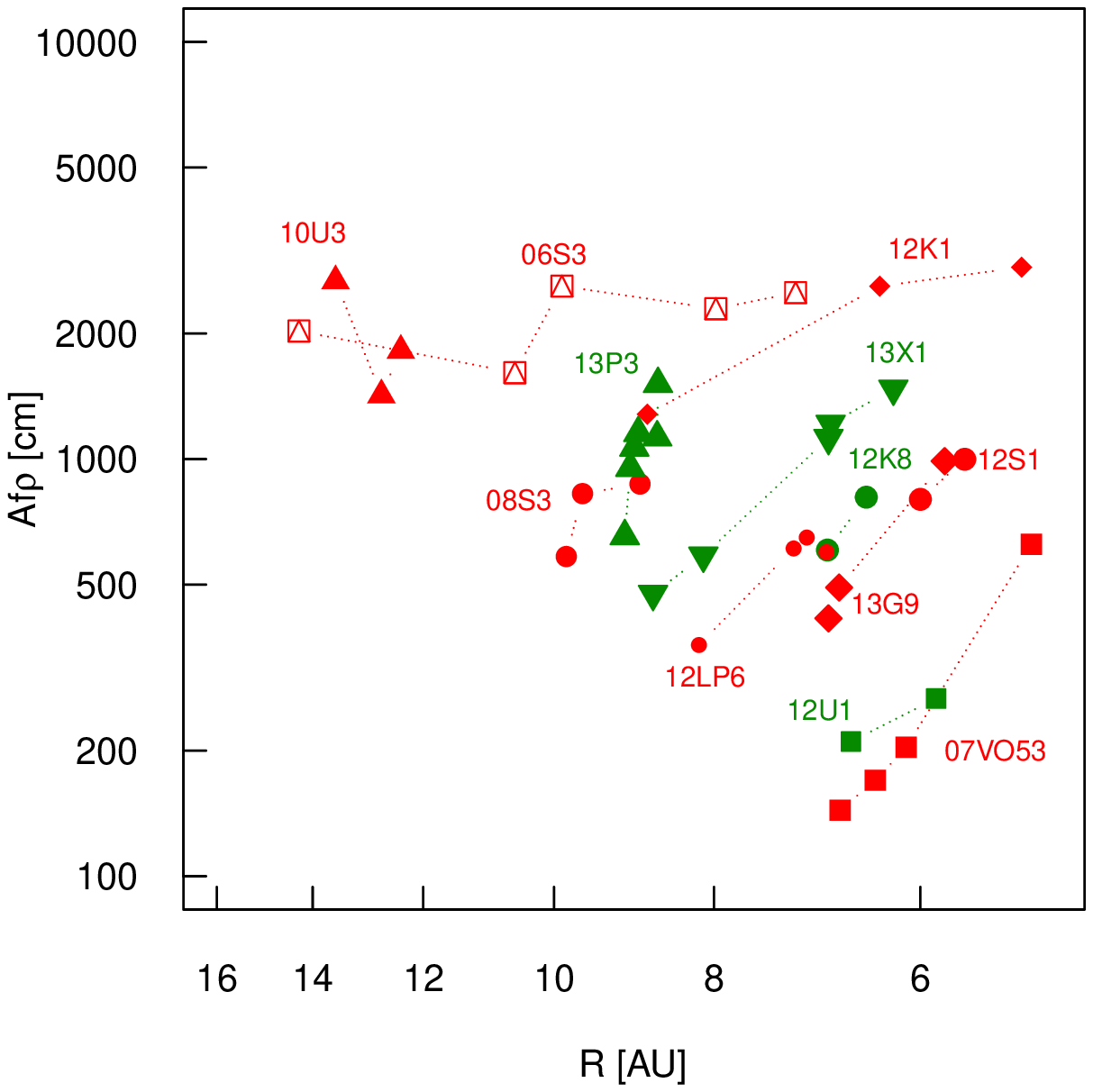}
\hfill\includegraphics[angle=270,width=8.0cm]{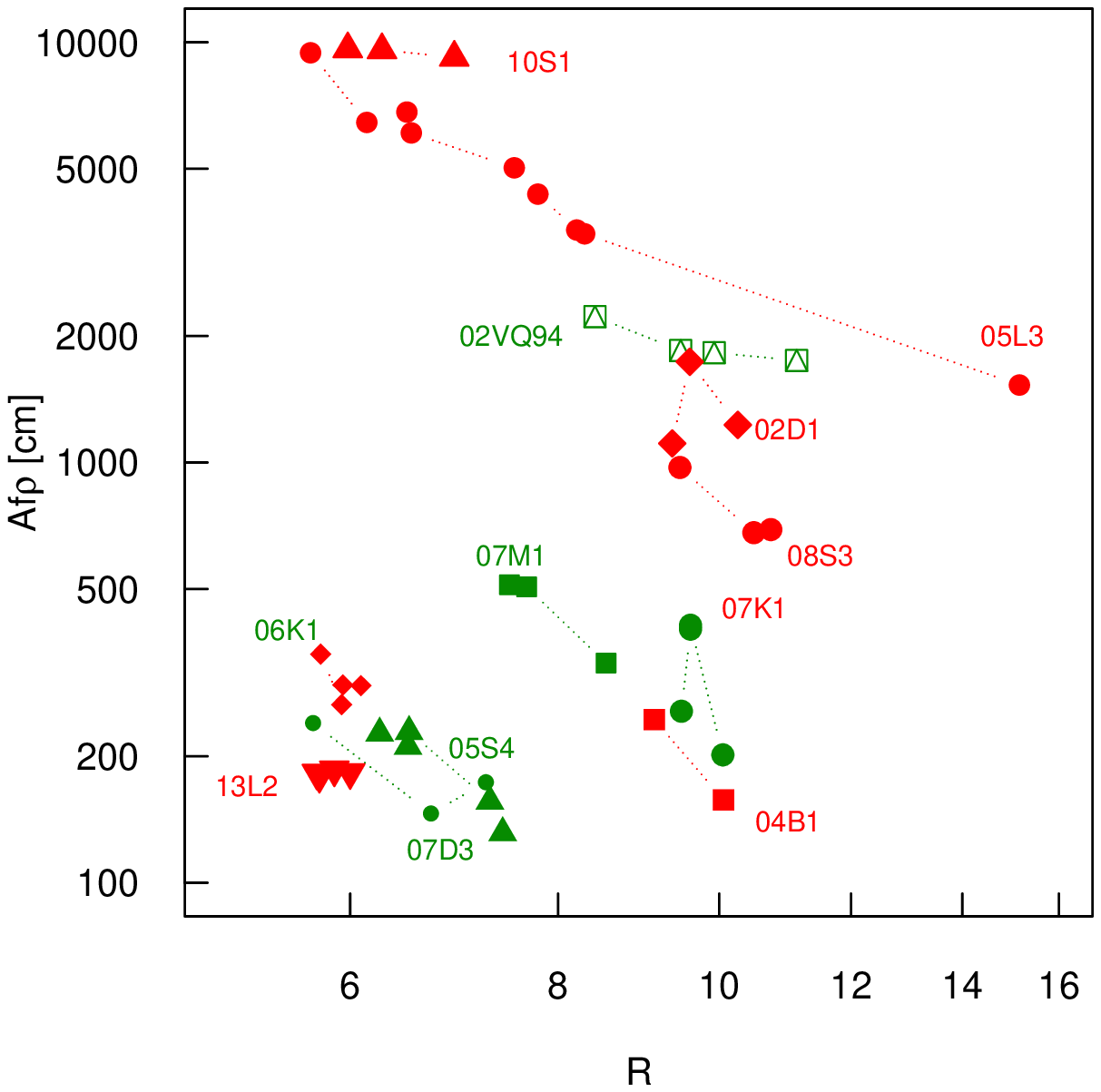}
\caption{Dependence of $Af\rho$ on the heliocentric distance (log-log scale). The left panel shows comets pre-perihelion, right panel plots the same for post-perihelion. Note that the abscissa in the left panel decreases toward right, so the Sun can be considered in between the two panels.}
\end{figure*}

\bigskip
\noindent
{\it C/2007 D3 (LINEAR)}
\newline
\noindent

This returning comet ($P\approx 21000$ years) was discovered by the LINEAR team
on February 20, 2007 at a heliocentric distance of 5.2 AU, several months prior
to perihelion pssage.  We observed this comet on the outward orbit between
5.70 AU (2008-03-31) and 7.24 AU (2009-04-19).  Ivanova et al (2015) observed
this comet about two weeks prior our fist observation, they derived derived an Afrho about 180-200 cm, which is consistent to our data.

Our observations covered a year. We detected a coma diminishing between 10--5\arcs. A dust tail was observed with lengths 10--80\arcs (2008-03-31: 40$^{\prime\prime}$, PA 240, wide shape; 2008-12-31: 80$^{\prime\prime}$, PA 250, narrow shape; 2009-04-18: 10$^{\prime\prime}$, PA 225).

\bigskip
\noindent
{\it C/2007 M1 (McNaught)}
\newline
\noindent
This returning comet (P$\approx$51,000 years) was discovered again by R. McNaught, who identified the comet in June 2007 at 7.9 AU heliocentric distance, about 14 months prior the perihelion (7.5 AU). We observed this comet on the outward orbit, between 7.48 and 8.45 AU. Mazzotta Epifani et al. (2014) derived an $Af\rho_{max}$ = 484$\pm$8 cm at R = 7.69 AU. Due to the low S/N ratio, they did not describe a detailed morphological analysis, however, their image suggests a detection of an elongated coma without a tail.

Our results are compatible to the findings of Mazzotta Epifani et al. (2014). During 2008--2010, all images showed a diffuse coma without a tail, with $Af\rho$ values decreasing between 581--333 cm on the outward orbit.

\bigskip
\noindent
{\it C/2007 VO53 (Spacewatch)}
\newline
\noindent
The object was discovered as an asteroid by the Spacewatch project at large
heliocentric distance (8.3 AU) on Nov. 1, 2007. A faint coma was detected two and half
months later with the 1.8-m Spacewatch II Telescope. The orbital calculations suggested that this is a dynamically new comet with perihelion
distance of 4.84 AU, a perihelion passage was on April 2010. The maximum brightness was around 17$^{\rm m}$. 

We observed this comet on the inward orbit, between 6.70 and 5.14 AU. Interestingly, this was among the faintest comets in our survey, and it exhibited the lowest $Af\rho$ level of 143 cm at 6.71 AU heliocentric distance on 2008--08--21. That time the comet was star-like, suggesting a faint, very compact coma. By 2008 December, however, the view dramatically changed: an expressed coma appeared, and a short faint tail developed toward PA 170. On 2009--09--16, the general appearance was similar, but the tail was even expressed and longer, reaching a length of 60$^{\prime\prime}$ at more least. Evidently, this comet exhibited the most spectacular brightening and morphological evolution on its inward path.

\begin{figure*}
\includegraphics[angle=270,width=8.0cm]{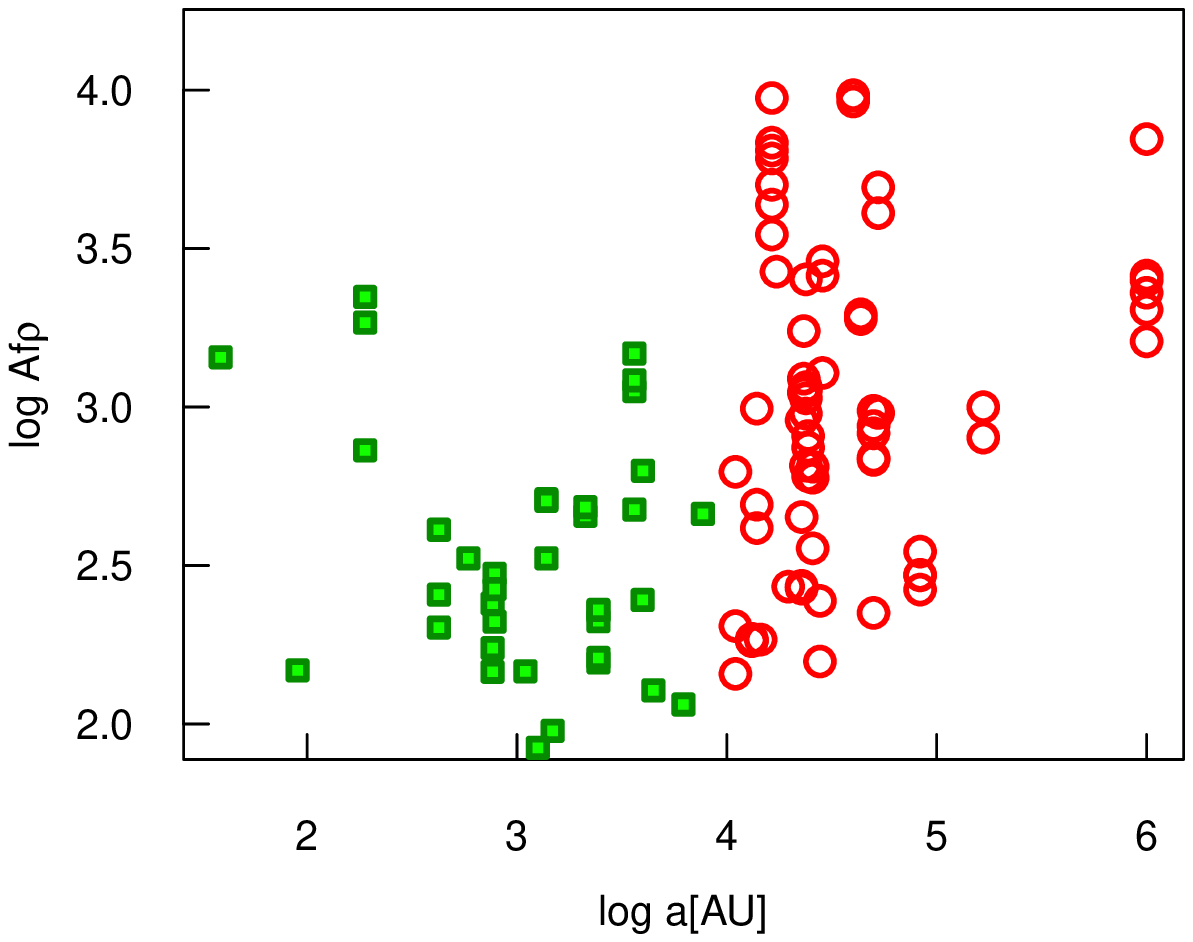}
\hfill\includegraphics[angle=270,width=8.0cm]{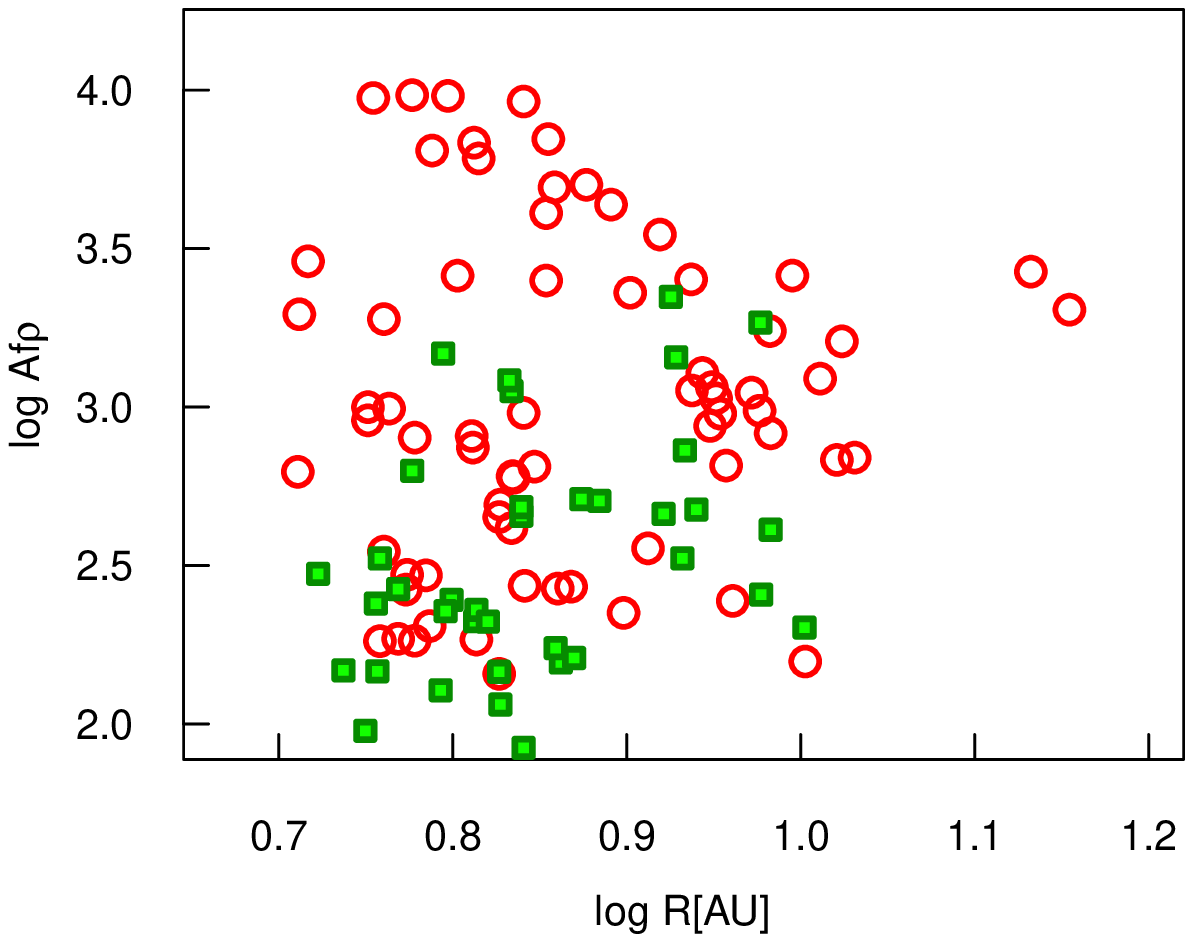}
\caption{Left panel: dependence of \afrho on the semi-major axis. Recurrent comets are denoted by green boxes, while Oort-cloud comets are plotted with red open circles. Right panel: the same but as a function of the heliocentric distance at observation.}
\end{figure*}

\bigskip
\noindent
{\it C/2008 S3 (Boattini)}
\newline
\noindent
The dynamically new comet was discovered by Andrea Boattini in the CSS/MLS project in 
Sept. 2008 at 10.0 AU heliocentric distance. Further prediscovery images were found on CSS/MLS frames from Dec. 2007 and Dec. 2006 at 10.9 and 12.4 AU at respectively. The perihelion passage occurred at 8.02 AU on 2011 June, maximum CCD brightness of 17\fm5. The derived absolute brightness is about H$_{10} = 4$\fm0 mag.

This comet was observed on both the inward (8.87--9.83) and outward (9.47--10.74) orbits, and gives a prime opportunity to compare the inward/outward behaviour of a dinamicaly new comet. The comet continuously exhibited a compact but not star-like coma, and a faint tail with a length of 20--30$^{\prime\prime}$. The visibility of the tail followed the changes of  $Af\rho$; while its shape underwent some evolution on the orbit. Before perihelion, the tail was narrower, and after perihelion, it turned into a more extended, fan-like feature, with its bisector pointing toward the predicted direction of the dust tail.

The general conclusion is that the activity is significantly higher post-perihelion than on the pre-perihelion orbit. This kind of behaviour is similar to that observed for C/2006 S3, and strengthens the general conclusion that new comets tend to be more active post-perihelion.

\bigskip
\noindent
{\it C/2010 R1 (LINEAR)}
\newline
\noindent

The dynamically new comet was discovered by the LINEAR team on September 4,
2010 at 7.2 AU heliocentric distance, about 20 months prior the perihelion
(5.6 AU). We observed this comet on the inward orbit at 5.64 AU (2012-03-18) and on
the outward orbit at 7.25 AU. Mazzotta Epifani et al. (2014) derived an
$Af\rho703\pm56$~cm at $R = 6.09$~AU on the inward orbit.

On 2012-03-17 an outer coma of 10\arcs was observed with a 20\arcs long tail toward PA 105. By 2014-02-23, the coma reduced to 5\arcs diameter, while a narrow, very long tail was observed, reaching 130\arcs length.

\bigskip
\noindent
{\it C/2010 S1 (LINEAR)}
\newline
\noindent
The comet is a dynamically new which was discovered on Sept. 2010 by the LINEAR team as an asteroid-like object. Follow-up observations revealed a compact coma and faint tail. The comet passed its perihelion on May 20, 2013 at a distance of 5.90 AU from the Sun and a visual brightness of 13\fm0.  The derived absolute brightness is about H$_{10} = 3$\fm0 mag. 

Recently, Ivanova et al. (2015) derived an 3900~cm, at 7.00 AU distance, and Shubina et al (2014) published an $Af\rho$ of 8400$\pm$600 cm in $V$ and 8200$\pm$1000 cm in $R$ band, at 6.33 AU distance on the inward orbit, respectivelly.

We observed this comet on the outward orbit, between 5.91 and 6.93 AU, and detected the decreasing trend of $Af\rho$ between 9600--9200 cm, which is consistent with the Shubina et al. (2014) pre-perihelion data.

Several months after perihelion, the comet showed a large, diffuse coma (1$^\prime$ in diameter) with a strong central condensation. A prominent, 3--4$^\prime$ long tail developed toward $\approx$PA~40 deg. In 2014 autumn, the tail had a similar size but the tail generally dimmed. That time, the most characteristic morphological feature was a fan-shaped, bright outflow extending to 25$^{\prime\prime}$ toward PA 220$^\circ$, where after a complete reversal, its material turned back toward the tail (PA 40$^\circ$).

\bigskip
\noindent
{\it C/2010 U3 (Boattini)}
\newline
\noindent
The dynamically new comet was discovered by Andrea Boattini in the CSS/MLS project in 2010 October. Its heliocentric distance (18.4 AU) has been the largest discovery distance of a comet ever, and this time the comet had more than eight years to the perihelion passage. The comet is still far away from perihelion on the inward orbit, as the passage will occur at 8.45 AU in 2019 February. The derived absolute magnitude is about H$_{10} = 1$\fm0 mag, the brightest in our sample.

In 2013 December the comet showed slightly diffuse nucleus of 4$^{\prime\prime}$ diameter in the southern end of an elliptical coma
in 8$\times$15$^{\prime\prime}$ diameter. In the second half of 2014 the comet was fainter than the first observation, and the circular coma was diffuse and ill defined. This change of appearance and the decreasing $Af\rho$ suggested that the comet was undergoing an outburst in 2013 or before.

\bigskip
\noindent
{\it C/2012 K1 (PANSTARRS)}
\newline
\noindent

This retrograde Oort cloud comet was discovered by the Pan-STARRS project on
May 19, 2012. at 8.8 AU heliocentric distance. The object shows condensed coma
with diameter of several arcseconds, with total magnitude about 19.5. The
comet came to perihelion on 27 August 2014 at a distance of 1.05 AU from the
Sun. We observed this comet three times on the inward orbit at 8.78, 6.35 and
5.21 AU, respetively.

On 2012-05-20 and 2013-03-12, the coma was compact, while a 10\arcs long tail has appeared by 2013 March. On 2013-07-14, the coma was quite large (20\arcs) and the tail reached 60\arcs (PA 120).

\bigskip
\noindent
{\it C/2012 LP26 (Palomar)}
\newline
\noindent
The dynamically new comet was discovered as an asteroidal object by the Palomar
Transient Factory program on June 10, 2012 at a heliocentric distance of 9.9
AU. A faint coma was detected eight months later with the 0.9m Spacewatch
Telescope, at 8.9 AU heliocentric distance. The comet reached perihelion more
than three years after the discovery (2015-08-16) at 6.53 AU.  We observed
this comet four times on the inward orbit between 8.17 and 6.84 AU.

Both the extension and the brightneess of the coma increased during the one year of our observations. The coma was 5\arcs on 2013-08-08 and 2014-06-03, while a 10\arcs (PA 265) long tail also appeared by 2014 April. By 2014 July, the coma increased to 8\arcs and the tail to 15\arcs (PA 240).

\bigskip
\noindent
{\it C/2012 U1 (PANSTARRS)}
\newline
\noindent

This large perihelion distance (5.26 AU) returning comet ($P\approx 32.000$~years) was discovered by the Pan-STARRS project on October 18, 2012 at a
heliocentric distance of 7.0 AU, more than 20 month prior to perihelion
passage. We observed this comet four times on the inward orbit between 6.61 and
5.87 AU, the first and second pairs of observations were separated by only several 
days. Our fifth observation we made on outward orbit, fifty-two days after
perihelion at 5.28 AU. 

This comet was characterized by a diffuse coma (5\arcs) at the time of our observations in 2013. On 2014-08-25, the coma increased to 10\arcs, the tail appeared with 25\arcs length toward PA 300.

\bigskip
\noindent
{\it C/2013 G9 (Tenagra)}
\newline
\noindent
 
This apparently asteroidal object was reported by M. Schwartz and P. R. Holvorcem
on images obtained with the Tenagra II astrograph on April 15, 2013. A compact
coma was detected several days later at 7.1 AU. The orbital calculations
suggested that this was a dynamically new comet with perihelion distance of
5.34 AU, a perihelion passage was on January 14, 2015. We observed this comet
three times on the inward orbit at 6.83, 6.72 and 5.80 AU, respetively.

The comet exhibited a similar picture in all images, as a condensed coma completed by a short (5\arcs) tail. By 2014-03-13, the tail increased to 10\arcs.

\bigskip
\noindent
{\it C/2013 X1 (PANSTARRS)}
\newline
\noindent
This dynamically new comet was discovered using the 1.8-m Pan-STARRS telescope
in Dec. 2013, when it was still 8.9 AU from the Sun with an apparent magnitude of 20. The perihelion passage will be on April 20, 2016 at heliocentric distance of 1.31 AU, when this comet might become a bright, 7--8 magnitude object. We observed this comet on the inward orbit, between 8.72 and 6.26 AU.

The comet showed stellar appearance at the end of 2013, and became a small fuzzy object with the same brightness
in spring of 2014. After the conjuction the coma remained fuzzy, but a faint short tail evolved toward PA 30$^\circ$
(2014--08--25). In autumn of 2014 comet showed a strange,
rectangular appearance with a coma$+$tail approximately
9x15$^{\prime\prime}$ dimension. During this path, $Af\rho$ increased monotonically between 474--1474 cm.

\section{Discussion}

\begin{figure*}
\includegraphics[angle=270,width=8.0cm]{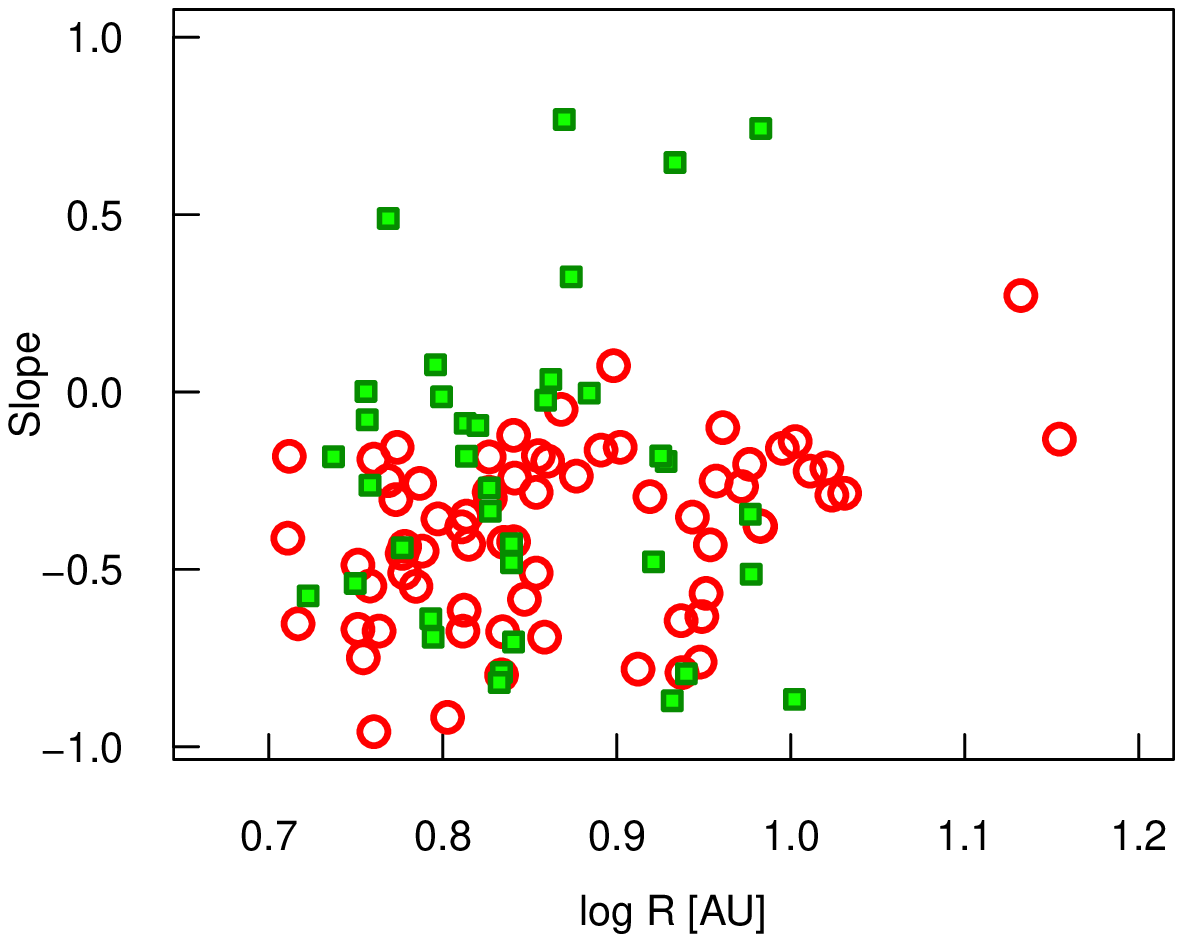}
\hfill\includegraphics[angle=270,width=8.0cm]{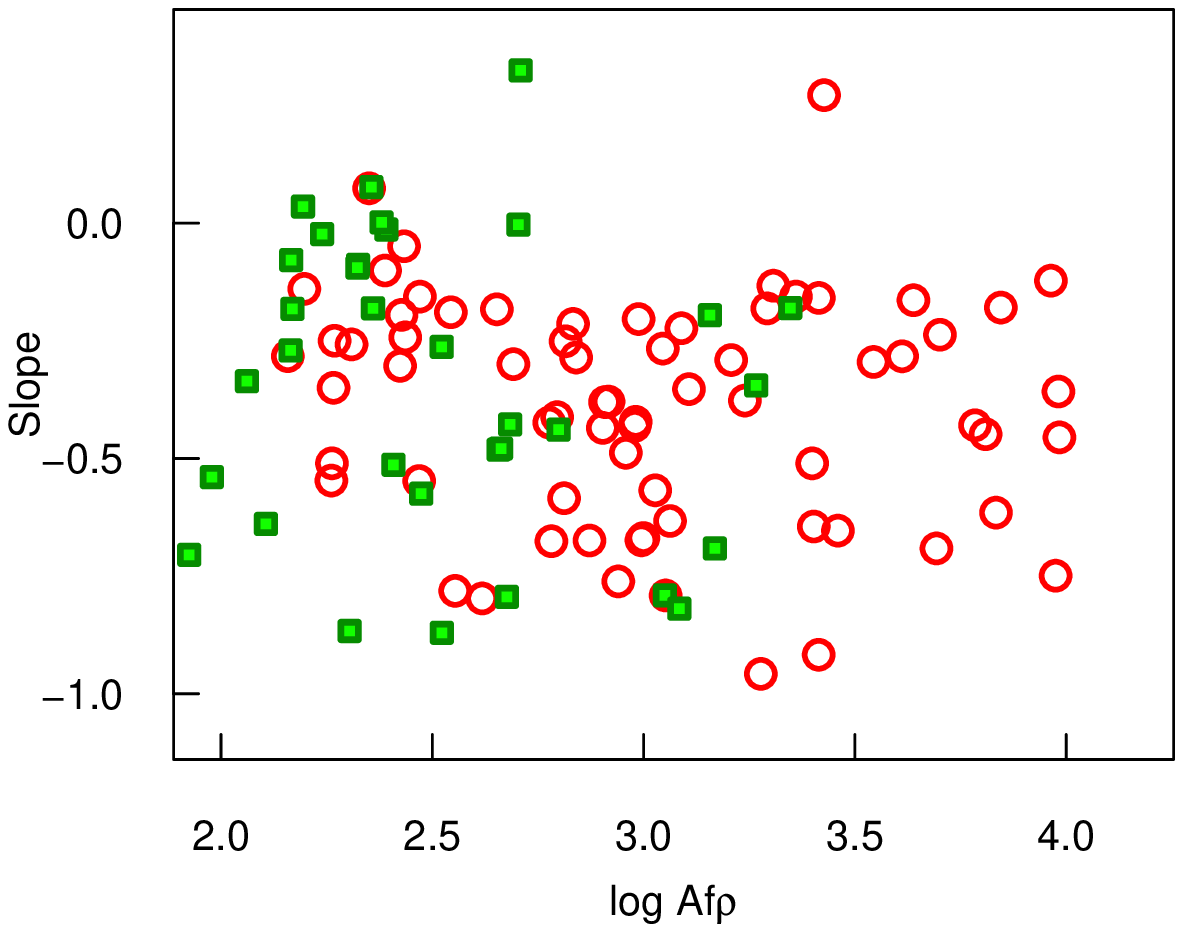}
\caption{Left panel: Dependence of the slope on the heliocentric radius. Right panel: the same as a function of \afrho. The color coding is the same as in Fig. 3.}
\end{figure*}

The observation program we report here looks back to more than 15 years. There are a few comets which have been observed both before and after the perihelion with long coverage (most importantly C/2008 S3, C/2002 VQ94, C/2006 S3). The systematic and comparative study of the activity of LP comets beyond 5 AU requires observations covering at least 15-25 years, most importantly because requires systematic observations of comets which are discovered well before the perihelion. The work we present here can be considered as a basis for such a survey, since it already contains C/2008 S3 with a long inward$+$outward observation history (and a few comets with few points inward and outwards, e.g. C/2002 VQ94, C/2006 S3).


About ten comets in the presented sample are worth further follow-up lasting for many years, or even decades. These comets are typically around the perihelion, or somewhat after it now, have a well covered inward history in our data, and following the outward activity is highly desirable (C/2006~S3, C/2008~S3, C/2010~S1, C/2012~K8, C/2012~LP26, C/2012~U1, C/2013~G9, C/2013~P3). The most promising objects for further observations are those ones which are still before perihelion. C/2010~U3, being a unique comet for its early discovery, will reach the perihelion only in 2019. C/2013~X1 will have a very close perihelion in 2016, approaching the Sun to 1.3 AU. C/2011~KP36 and C/2010~U3 will reach their perihelion also in 2016. The brightest targets (C/2005~L3, C/2006~S3, C/2010~S1, C/2010~U3), where the size of the nucleus can be expected between those of comets Halley (Hainaut et al. 2004) and Hale--Bopp (Szab\'o et al. 2012), offer a good detectability of the naked nucleus after the cessation of the activity.

A visual inspection suggests that the appearance of the observed comets is rather diverse. With \afrho~values spanning over 2 orders of magnitude (indicatively 100--10,000 cm), there are comets which are compact or diffuse -- apparently mostly regardless to the brightness or the \afrho. A surprisingly large fraction of the comets have tails, which are quite impressive in some cases (see Fig. 1 for the best examples), and while these tails are rather tenious, they were observed around fainter comets as well, and they were not necessarily associated with the brightest comets. These visual findings can be quantitatively analysed by means of \afrho and slope parameters. We will evaluate the results from the three most intriguing aspects in the followings.

\subsection{The \afrho{}--$R$ activity profiles}

The evolution of \afrho with $R$ is plotted in Fig. 2, left and right panels. The left panel shows the pre-perihelion activity, while the right panel shows the same, but for the post-perihelion orbit. The heliocentric distance in the left panel increases leftwards, therefore we can imagine that the Sun is at the center, between the two panels. The symbol and color coding discriminate the recurrent comets (green boxes) and Oort-cloud comets (open red circles). 

 The general $\log$\afrho--$\log R$ profile is roughly linear for the most comets (Fig. 2), implying an underlying power law, which we can confirm for a large set of LP comets. Interestingly, the rate of increasing of the activity seems to depend on the onset level of the baseline activity level. Comets which exhibit high \afrho values far from the Sun tend to follow a shallower activity history, while other comets which are relatively fainter/less active at large solar distances can evolve more spectacularly, following a steeper power index. This observation is also compatible with the remarkable comet disappointments from the previous decades, when impressive comets, discovered at large solar distances with extraordinary activity level tend to slowly evolve and significantly underachieved the peak brightness predicted from early data (C/1973~E1 (Kohoutek), C/1989~X1 (Austin), C/2001~Q4 (NEAT)). Our dataset suggests that this is a usual pattern of LP comets which exhibit distant pre-perihelion activity of an extraordinary level. 

A similar behaviour is suggested for the post-perihelion phase, but may be with a less expresed difference between more and less active comets, while the more active ones seem to exhaust slowlier. The slow fading is a well known observation also for periodic comets, and is usually explained by the heat inertia of the nucleus.

\subsection{Comparison of dynamically new and recurrent comets}

Based on their database of comet observations ($\sim$50 comets over a range of 1 to $\sim$30 AU), Meech \& Hainaut (1997) have shown that dynamically young comets are intrinsically brighter, exhibiting dust comae and activity at large distances in the region where water ice sublimation is not possible. The observations by date support this tendency in general, and our results also support this conclusion.

In the two panels of Figures 3 and 4, we plot the dependence of \afrho on various orbital and morphological parameters (slope, heliocentric distance, semi-major axis). Note that there are more comets in these figures as in Fig. 2, since here we plot those comets which are represented by only one or two observations in the current dataset.  The cloud of points representing dynamically new comets (Fig 3, right panel, red points) lies convincingly above that of recurrent comets (green dots in the same figure panel), regardless to the solar radius. The difference seems to be most prominent between 5--7 AU, where the dynamically new comets tend to advance up to a factor of $\approx 4$ higher limit in \afrho, in comparison to recurrent comets. This observation is compatible to the presence of a huge amount of volatiles on the surface of new comets, in respect to returning comets.

The slope parameter also shows an interesting difference between returning and new comets (Fig. 4). Both groups tends to exhibit slightly negative slope, typically between 0 and $-1$, and also, both groups have comets which were observed at a non-steady state, with a slightly positive slope value. The difference between the two groups is that the observations of negative slope usually represents the returning comets in the sample. Several features can lead to positive slopes, such as fan-like or spinning features in the coma, and the sudden and usualy temporal decrease of the activity when the outer coma is relatively richer in dust, in respect to the inner coma. The increased occurrence of positive slope in the returning group reflects that such uncommon morphological features and/or puffing activity is more common in the returning group, while activity of the new comets is usualy more regular.

It is hard to say wether there were sudden changes in the activity level or morphological parameters during the observed period of the target comets, simply because there is not enough dense coverage for continuous monitoring of such effects. However, several comets have observations separated at 1-2-3 days apart (see Table 2), and during these observations, the observed parameters did not change exceeding the observational errors. Therefore, we have no evidence of sudden events in the observed data series. In essence, the most important groupping of the dataset is related to the recurrent/new nature of the comet. Dynamically new comets are characterized by a higher level of activity in average, with more regular and smoothly evolving matter production. The more symmetrical appearance suggests also a more isotropic outflow, too.

\section{Discussion -- ways to further interpretation}

The observed diverse behaviour in our sample is difficult to interpret simply because of the lack of more information. The spacecraft missions have revealed that comets span over a wide spectrum of different surface structures and morphology. These missions revealed a very little amount of volatiles of the surface. This latter observation is also not too informative for the recurrent and dynamically new comets, since they can host a more significant surface volatile content on their more intact surface. The thermophysical factors are the real bottleneck of modelling the activity: observations by the Herschel Space Observatory have shown that the surface temperature maps of asteroids are very complex. Due to their general similarity, the same must happen on cometary surfaces, too, instead a homogenuous surface -- which is an assumption of most cometary activity models -- there are likely steap temperature gradients both laterally and radially, and also similarly to the asteroids, the depth of the diurnal layers is also likely to vary significantly with the position.  The extreme temperature inhomogeneities with diurnal and probably seasonal variations were in fact observed by the ROSETTA mission on 67P/Churyumov-Gerasimenko (e.g. Gulkis et al. 2015, H\"assig et al. 2015). For long period comets, another drawback is that we do not even have a confirmed observation of the naked nucleus of a long period comet, therefore the most important variable in the activity models -- the size of the nucleus -- is still unknown. Since the local structures are intimately related to the general shape, which is unknown, any modeling effort can be very draft. 

In this aspect, the forthcomming sky surveys, most importantly the LSST, can significantly help. LSST will be capable to observe the dormant nuclei of comets up to 20-50 AU (indicatively in the size range of 10--60 km radius). These sample will consist of several currently active long period comets, which will be dormant by the time of LSST observations; and also, pre-perihelion dormant comets that will be probably years later discovered as active comets, when they will get closer to the Sun. Another aspect of LSST will be its capability to observe the activity of the long period comets during the entire visibility with an unprecedentedly dense cadence, allowing the comparison of pre- and post-perihelion activity, and the quest for short period variations for several objects.

However, this latter task will be difficult to complete even by LSST. Our current survey shows that this 15 year coverage is still insufficient for covering both the inward and outward activity of the same comet, simply because distant comets move slowly. In the foreseeable future, this task will be still mostly devoted to specific mini-surveys, like the one for which we presented the early years in this paper.




\section{Summary}

Our results can be summarized as follow. 

\begin{enumerate}
\item We present 152 observations of 50 comets from 103 nights, showing activity at solar distances between 5--15 AU. \afrho{}, \afrhomax{} and slope parameters were determined for all observations.

\item{} All comets showed a coma and in many cases a tail, which sometimes exceeded 10$^\prime$. 

\item{} The \afrho{} value of the most active recurrent comet was surpassed by that of 5 new comets. The average value of \afrho{} of recurrent comets significantly exceeded that of the recurrent comets. We explained this observation with an increased amounts of volatiles on the surface of the new comets, demonstrating that dynamically new comets are also new in composition.

\item{} All new comets showed a negative slope parameter and usualy a quite symmetric coma. The observed positive slope parameters, that we interpreted as signs of activity variations, or jets, arcs and other internal structures, were observed only in recurrent comets.

\item{} The evolution of \afrho{} roughly followed a power law of $R$ for the most comets. Those comets which were characterized by larger \afrho{} values at larger solar distances tended to follow a flatter power. This suggested that comets which had very significant activity around 10--15 AU tended to less increase in activity with decreasing solar distance, thus representing the ''comet disappointment'' scenario.

\item{} The analogous ``inverse'' comet disappointment behavior was observed post-perihelion, but with a less expressed difference between the most and least active comets.

\item{} We examined the dependence of the studied activity parameters on orbital elements, the actual ephemerides at the observations and the visibility of a tail. We did not find significant subgroups, the only significant grupping in the sample was the new/recurrent type of the comet.

\item{} Since the fully covered observation of a comet, extending beyond 10--15 AU solar distances on both half orbits lasts decades, we proposed comet observations with very long time coverage of 20--25 years, to be able to compare the inward and outward behavior of the same comets in an extensive sample.

\end{enumerate}

\acknowledgements
We acknowledge useful discussions about comet activities with P.~Cs. Kiss. We thank E. B\'anyai, G. Hodos\'an, \'A. K\'arp\'ati, Z.  Kuli, Sz. M\'esz\'aros, Gy. Mez\H o, A. Nagy, A. E. Simon, B. Sip\H ocz, A. Szing and K. Tak\'ats for some observations on our long-time comet research program. We also acknowledge the Hungarian OTKA Grants K-104607, K-109276 and K-113117, the ESA PECS Contract No. 400011{}\-{}0889/14/NL/NDe, the Lend\"ulet-2009 Young Researchers Program and the LP2012-31 grant of the Hungarian Academy of Sciences and by the City of Szombathely under agreement No. 61.360-22/2013.

\end{document}